\documentclass[twocolumn,showpacs,preprintnumbers,prd,nofootinbib]{revtex4-1}
\usepackage{amsmath}
\usepackage{amsfonts}
\usepackage{amssymb}
\usepackage{graphicx}

\begin{document}

\title{A scalar field dark matter model and its role in the large scale structure 
formation in the universe}
\author{Mario A. Rodr\'\i guez-Meza}
\affiliation{
Departamento de F\'{\i}sica, Instituto Nacional de
Investigaciones Nucleares, 
Apdo. Postal 18-1027, M\'{e}xico D.F. 11801,
M\'{e}xico. e-mail:
marioalberto.rodriguez@inin.gob.mx
}

%\authorrunning{M.A. Rodr\'\i guez-Meza}

%\thispagestyle{empty}
%\maketitle
%\thispagestyle{empty}
%\setcounter{page}{87}

\begin{abstract}
%\begin{abstract}
In this work we present a model of dark matter based on scalar-tensor theory of gravity.
With this scalar field dark matter model we
study the non-linear evolution of the large scale structures in the universe.
%using a dark matter
%model steaming from a scalar-tensor theory of gravitation.
The equations that govern the evolution of the scale factor of the universe are derived
together with
%and also 
the appropriate Newtonian equations to follow the non-linear evolution
of the structures. 
Results are given in terms of the power spectrum that gives quantitative
information on the large-scale structure formation. 
The initial conditions we have used are consistent
with the so called concordance $\Lambda$CDM model. 
%We also show how to estimate
%the scalar field dark matter parameters using a sample of observed rotation curves.
%\end{abstract}
\end{abstract}

\maketitle

%%%%%%%%%%%%%%%%%% INTRO %%%%%%%%%%%%%%%%%%
\section{Introduction}\label{sec:Intro}

%The Big Bang theory 
The standard model of cosmology
is supported by three main astronomical observations: 
the surveys of supernovae Ia, the cosmic microwave background radiation (CMB), and
the primordial nucleosynthesis. These observations together with other
%
%M
modern cosmological observations, 
like galaxies surveys (SDSS, 2dF), galaxy rotation curves, the Bullet Cluster observation, studies
of clusters of galaxies,
establish that the universe behaves as dominated
by dark matter (DM) and dark energy. However, the direct evidence for the existence of  these
invisible components remains lacking. Several theories that would modify our understanding
of gravity have been proposed in order to explain the large scale structure formation in the 
universe
and the galactic dynamics. The best model we have to explain the observations is the 
$\Lambda$CDM model, i.e., the model of cold dark matter (CDM) 
--non-relativistic particles 
%at the epoch of last scattering 
of unknown origin-- with cosmological constant 
($\Lambda $), in particular,
this model explains very well the universe on scales of 
galaxy clusters and up \citep{Breton2004}. 

The $\Lambda$CDM model has become the theoretical paradigm leading
the models of the universe to explain the large scale structure (LSS)
 formation  and several other
observations.
Where ``large'' means 
scales larger 
than $1$ Mpc --about the size of the group of galaxies that our Milky Way belongs.
Together with the cosmic inflation theory, this model makes a clear prediction about
the necessary initial  conditions that the universe has to have in order to have the
structures we observe and that those structures build hierarchically due to a gravitational
instability.
One of its main predictions is that the density profile of galaxies, clusters of galaxies,
and so on, is of the form \citep{Navarro96,Navarro97},
\begin{displaymath}
\rho_{NFW}(r) = \frac{\rho_0}{(r/r_0)(1+r/r_0)^2},
\end{displaymath}
a density profile known as Navarro-Frenk-White profile (NFW).
Parameters $\rho_0$ and $r_0$ must be fitted, for example, using rotation
curves of galaxies.

The $\Lambda$CDM model and its 
success in explaining several  observations --this is why this model is also known
as the concordance model--
 yields the following conclusions:
 On large scales, the universe is homogeneous and isotropic, as described by the
 Friedmann-Lama\^itre-Robertson-Walker (FLRW) metric.
The geometry of the universe is flat, as predicted by inflation.
The dark matter is cold (non-relativistic at decoupling epoch).
The initial density fluctuations were small and described by a Gaussian random field.
The initial power spectrum of the density fluctuations was approximately the Harrison-Zeldovich spectrum ($P(k) \propto k^n$,   $n=1$) \citep{Zeldovich1970, Binney}.

In terms of the composition of the universe, the above conclusions can be summarized 
as follows:
Hubble's constant
(Expansion rate of the universe at the present epoch): 		
$H_0=73.2\pm 3.1$ km/s/Mpc.
Density parameter 
%of density
(combined mass density of all kind of mass and energy
in the universe, divided by the critical density):
$\Omega_0=1.02\pm 0.02$.
Matter density parameter
 (combined mass density of all forms of matter in the
universe, divided by the critical density):
$\Omega_m=0.241\pm 0.034$.
Ordinary matter parameter density
(density of mass of ordinary atomic matter
in the universe divided by the critical density):
$\Omega_b=0.0416\pm 0.001$.
Density parameter of dark energy
(energy density of dark energy in the universe
divided by the critical density):
$\Omega_\Lambda=0.759\pm 0.034$ \citep{Komatsu2010}.

Even though of all successes of the 
$\Lambda$CDM,
this model has several problems. Some of them are:
The exotic weakly interacting particles proposed as dark matter particles candidates
are still undetected in the laboratory.
The number of satellites in a galaxy such as the Milky Way is predicted to be an order of magnitude larger than is observed. 
 Cuspy halo density profiles.
%
%  The number of superclusters observed in surveys of galaxies (like SDSS) data 
%  appears to be an order of magnitude larger than predicted by $\Lambda$CDM simulations.
%
  The lack of evidence in the Milky Way for a major merger is hard to reconcile with the amount of accretion  predicted by $\Lambda$CDM.
% 
%The missing matter problem in the Coma cluster is a factor of  the order of 100 less than
%when Zwicky first discovered.
%
With respect to the inclusion of the cosmological constant, the ratio of the vacuum energy
density to the radiation energy density after inflation is 1 part in $10^{100}$, a fine
tuning coincidence. The cosmological constant has the wrong sing according to the
string theorists who prefer a negative $\Lambda$ instead of a positive $\Lambda$. 
The $\Lambda$CDM model predicts that large structures should form last and therefore should be
young whereas observations tell us that the largest galaxies and clusters appear old.
%
% Poner aqu' la discusi—n sobre los trabajos de Saridakis y Barrow
Several candidates have been proposed in the past that pretend to substitute the role that the cosmological
constant plays to accelerate the expansion of the universe. That the universe is expanding is supported mainly by
the observations of the supernova project SNIa\cite{Riess1998,Perlmutter1999}. And we have lead to conclude that
the universe is now dominated by an energy density with negative pressure and occupies about 70\% of the universe in the
present epoch. This energy is called generically the 
dark energy. Several models to explain this dark component has been proposed, and they may be classified 
accordingly to its equations of state as the following:
quintessence dark energy\cite{Ratra1988,Wetterich1988}, phantom energy \cite{Caldwell2002,Caldwell2003},
or the quintom cosmology paradigm\cite{Feng2005} (see also the review\cite{Saridakis2010} for even more details).
Cosmological constant has as its equation of state one in which the pressure is the negative of the density.

From the $N$-body numerical simulations point of view we have two works, in which a detailed analysis of two scalar field possibilities has
been explored. One is the coupled quintessence models \cite{Barrow2011a} and the other is the extended quintessence 
models \cite{Barrow2011b}. The former considers a scalar field coupled minimally to the Ricci scalar and is coupled to 
the Lagrangian matter contribution. The later is of the type of a scalar-tensor theory in which the scalar field is introduced 
to model dark energy with a potential that is an inverse power law. The model we present in this work
 pretends to model the dark matter
contribution in the large scale structure formation with a scalar field model that stems from the newtonian limit of a general scalar-tensor theory \cite{Rodriguez2001}.

% BULLET CLUSTER
%Also,
Another problem that the $\Lambda$CDM problem is facing is the following. 
Almost a decade ago a bow shock in the merging cluster 1E0657-56, known as the 
Bullet Cluster, observed by satellite Chandra indicates that the subcluster --found by 
\cite{Barrena2002}-- moving through this massive ($10^{15} h^{-1}$ M$_\odot$)
main cluster creates a shock with a velocity as high as 4700 km s$^{-1}$
\citep{Markevitch2002, Markevitch2006}. A significant offset between the distribution of X-ray
emission and the mass distribution has been observed \citep{Clowe2004,Clowe2006},
also indicating a high-velocity merger with gas stripped by ram pressure. Several authors
have done detailed numerical non-cosmological simulations 
\citep{Takizawa2005, Takizawa2006,Milosavljevic2007,Springel2007,Mastropietro2008}.
One of the key input parameters for the simulations is to set the initial velocity of the subcluster,
which is usually given at somewhere near the virial radius of the main cluster.
\cite{Lee2010} have run cosmological N-body simulation using a large box (27 $h^{-3}$ 
Gpc$^3$) to calculate the distribution of infall velocities of subclusters around massive
main clusters. The infall velocity distribution were given at 1--3 $R_{200}$ --similar to 
the virial radius-- and thus it gives the distribution of realistic initial velocities of subclusters
just before collision. This distribution of infall velocities must be compared with the 
best initial velocity used by \cite{Mastropietro2008} of 3000 km s$^{-1}$ at about
$2R_{200}$ to be in agreement with observations. \cite{Lee2010} have found that
such a high infall velocity is incompatible with the prediction of the $\Lambda$CDM.
Therefore, there are plenty of problems that the concordance model have to solve and
 finally, it does not tell us what is dark matter and dark energy.

The program to study the large scale structure formation should be to start with primordial
initial conditions which means give the initial relevant fields, such as for example,
density and velocity fields at the epoch of last scattering ($\sim z=1100$, the value
of the redshift at that epoch, i.e., 
a photon emitted at that epoch is redshifted as $1+z=\lambda_r/\lambda_e$, with
$\lambda_e$ the wavelength of the photon when emitted, and $\lambda_r$ is
the wavelength of the same photon when observed;
 the expansion factor for a universe with a flat geometry
is related to the redshift as $1+z=1/a$). 
Then, evolve this initial condition
using an $N$-body
scheme up to the present epoch ($z=0$).

Some questions we have to answer are:
What are the distribution of the LSS sizes?
  What is the amount of mass and its distribution at large scales?
  How are the voids distributed through the space? Are these voids devoid of any matter?
 How the LSS evolve with time?
  What is the DM equation of state? What is its role in the  LSS formation
   processes and galactic dynamics?
 What are the implications of the observed LSS on the cosmological model of our universe? And on the structure formation? 
 And of course, what is the nature of the dark energy and matter?

During the last decades there have been several proposals to explain DM, for example: 
Massive Compact Halo Objects
(Machos), 
Weakly Interacting Massive Particles (WIMPs) 
such as
supersymmetric particle like the neutralino. Other models propose that
there is no dark matter and use 
general relativity with an appropriate equation of state. 
Or we can use 
scalar fields, minimally or non minimally coupled to the geometry.
%
% Quizas revisar aqu' el trabajo de Li ...

In this work we are mainly concern with the problem of dark matter and its consequences
on the large scale structure formation process.  
Our DM model is based on using a scalar field (SF) that
is coupled non-minimally 
%with 
to the metric through the
 Ricci scalar
%geometry 
in the Einstein field equations. 
A scalar field is the most simple field of nature. 
Nordstrom proposed a gravity theory by 1912, before Einstein \cite{Faraoni2004}.
Scalar fields have been around for so many years since pioneering work of Jordan, Brans,
and Dicke. Nowadays, they are considered as: (a) inflation mechanism; (b) the dark
matter component of galaxies; (c) the quintessence field to explain dark energy; and
so on. Therefore, is natural to consider dark matter models based on modifications of
Einstein's general relativity that include scalar fields.
%
% Li, Mota and Barrow
In this paper we will show results about the role this scalar field plays on the non-linear large scale structure
formation of the universe. In particular, we will show how the power spectrums predicted by this model
compare with the power spectrums predicted by $\Lambda$CDM and the ones that come from observations.

So we organize
our work in the following form: In the next section we present the general theory of a typical
scalar-tensor theory (STT), i. e., a theory that generalizes Einstein's general relativity by including
the contribution of a scalar field that couples non-minimally to the Ricci scalar. 
In section \ref{sec:Cosmos} we show
how the Friedmann equations becomes within a STT and present our model for the evolution
of the universe expansion factor $a(t)$. In section \ref{sec:Nbody} 
we present the $N$-body method which
we will use to obtain the evolution of the large scale structures. Our results for an initial
condition of the fields that is consistent with the observations 
%and 
%the estimation of the parameters of the scalar field dark matter model
are given 
in section \ref{sec:Results}. Finally, 
our conclusions are given in section \ref{sec:Conclusions}.

%%%%%%%%%%%% TEORIA ESCALAR TENSORIAL %%%%%%%%%%%%%
\section{General scalar-tensor theory and its Newtonian limit}\label{sec:STT}

The Lagrangian that gives us the Einstein equations of general relativity is
\begin{equation}
{\cal L} = \frac{\sqrt{-g}}{16\pi G} R
\end{equation}
The Einstein field equations that are obtained from the above Lagrangian, in  the
limit of small velocities as compared with the speed of light and small forces, limit
known as the Newtonian limit, give us the standard Newtonian potential due to
%a body of 
a point particle of
mass $m$
\begin{equation}
\Phi_N= -G\frac{m}{r} 
\end{equation}
where $G$ is the gravitational constant. What we intend to do in this work is to obtain
the consequences in the LSS formation processes when we rise the constant $G$ to a scalar
field, 
$1/G \rightarrow \phi$. But we go beyond this approach and include in the Lagrangian
two additional terms that depend on this field, a kinetic and potential terms. We will show,
in particular, that the Newtonian limit of this theory gives for the Newtonian potential due
to a mass $m$ \citep{mar2004}, 
\begin{equation} \label{eq:YukawaPot01}
\Phi_N= -G\frac{m}{r} \left( 1 + \alpha e^{-r/\lambda} \right)
\end{equation}
i. e., the standard Newtonian potential is modified by an additional term that has
the form of a Yukawa potential.

Then, we start with the Lagrangian of a general scalar-tensor theory
\begin{equation}\label{EqSTTLagrangian}
{\cal L} = \frac{\sqrt{-g}}{16\pi} \left[ -\phi R + \frac{\omega(\phi)}{\phi}
	(\partial \phi)^2 - V(\phi) \right] + {\cal L}_M(g_{\mu\nu}) \; .
\end{equation}
Here $g_{\mu\nu}$ is the metric,
${\cal L}_M(g_{\mu\nu})$ is the matter Lagrangian and $\omega(\phi)$ and
$V(\phi)$ are arbitrary functions of the scalar field. The fact that we have
a potential term $V(\phi)$ tells us that we are dealing with a massive scalar field.
Also, the first term in the brackets, $\phi R$, is the one that gives the name of
non-minimally coupled scalar field.

When we make the variations of the action, $S=\int d^4x\, {\cal L}$,
with respect to the metric and the scalar field we
obtain the
 Einstein field equations \cite{Faraoni2004}
% (Faraoni, 2004)
\begin{eqnarray}\hspace{-1em}
R_{\mu\nu} - \frac{1}{2} g_{\mu\nu} R &=& \frac{1}{\phi}
\left[ 8 \pi T_{\mu\nu} + \frac{1}{2} V g_{\mu\nu}
+ \frac{\omega}{\phi} \partial_\mu \phi \partial_\nu \phi
\right. \nonumber \\
&& \left. -\frac{1}{2} \frac{\omega}{\phi}(\partial \phi)^2 g_{\mu\nu}
+ \phi_{;\mu\nu} - g_{\mu\nu} \, \square \phi \frac{\mbox{}}{\mbox{}}
\right]  , \label{eq:EinsteinEqs}
\end{eqnarray}
for the metric $g_{\mu\nu}$ and for the massive SF $\phi$ we have
\begin{equation}\label{EqSTTPhi}
\square \phi  = \frac{1}{3+2\omega} \left[
	8\pi T -\omega' (\partial \phi)^2 +\phi V' - 2V \right] \, ,
\end{equation}
where $()' \equiv \frac{\partial }{\partial \phi}$. Here $T_{\mu\nu}$ is the energy-momentum
tensor with trace $T$, $\omega(\phi)$ and $V(\phi)$ are in general arbitrary functions that 
govern kinetic and potential contribution of the SF. 
If in Lagrangian (\ref{EqSTTLagrangian}) we set $V(\phi)=0$ we get the Bergmann-Wagoner
theory. If we further set $\omega(\phi)= \text{constant}$ the Jordan-Brans-Dicke theory
is recovered. The gravitational constant is now contained in $\phi$.
Also, the potential contribution, $V(\phi)$,
provides mass to the SF, denoted here by $m_{SF}$.
%

%%%%%%%%%%%% NEWTONIAN LIMIT OF STT %%%%%%%%%%%%%
\subsection{Newtonian limit of a STT}\label{sec:NewtonianLimitSTT}

The study of large-scale structure formation in the universe is greatly simplified by the fact that a
limiting approximation of general relativity, the Newtonian mechanics, applies in a region
small compared to the Hubble length $cH^{-1}$ ($cH_0^{-1}\approx 3000 h^{-1}$ Mpc, where 
$c$ is the speed of light, $H_0=100 h$ km/s/Mpc, 
is Hubble's constant and $h\approx (0.65-0.75)$), and large compared to the Schwarzschild
radii of any collapsed objects. The rest of the universe affect the region only through a tidal field.
The length scale $cH_0^{-1}$ is of the order of the largest scales currently accessible in 
cosmological observations and $H_0^{-1} \approx 10^{10}h^{-1}$ yr 
characterizes the evolutionary time scale of the universe \cite{Peebles1980}.

Therefore, in the present study, we need to consider the influence
of SF in the limit of a static STT, and then we need to describe
the theory in its Newtonian approximation, that is, where  gravity 
and the SF are weak (and time independent) and velocities of dark matter
particles are
non-relativistic.  
We expect to have small deviations of
the SF around the background field, defined here as
$\langle \phi \rangle$ and can be understood as the scalar field beyond all matter.
Accordingly we assume that the SF oscillates around the constant background field
\begin{displaymath}
\phi = \langle \phi \rangle + \bar{\phi}
\end{displaymath}
and
\begin{displaymath}
g_{\mu\nu} = \eta_{\mu\nu} + h_{\mu\nu},
\end{displaymath}
where $\eta_{\mu\nu}$ is the Minkowski metric.
Then,  Newtonian approximation
gives \cite{Pimentel1986,Salgado2002,Helbig1991,mar2004}
\begin{eqnarray}
R_{00} = \frac{1}{2} \nabla^2 h_{00} &=& \frac{G_N}{1+\alpha} 4\pi \rho
- \frac{1}{2} \nabla^2 \bar{\phi}  \; ,
\label{pares_eq_h00}\\
  \nabla^2 \bar{\phi} - m_{SF}^2 \bar{\phi} &=& - 8\pi \alpha\rho \; ,
\label{pares_eq_phibar}
\end{eqnarray}
we have set $\langle\phi\rangle=(1+\alpha)/G_N$ 
and $\alpha \equiv 1 / (3 + 2\omega)$.  In the above expansion we have set
the cosmological constant term equal to zero, since on small galactic
scales its influence should be negligible.  
However, at cosmological scales we do take into account the cosmological
constant contribution, see below.

Note that equation (\ref{pares_eq_h00}) can be
cast as a Poisson equation for $\psi \equiv (1/2) (h_{00} + \bar{\phi}/\langle\phi\rangle)$,
\begin{equation}\label{eq:Psi}
 \nabla^2 \psi = 4\pi \frac{G_N}{1+\alpha}  \rho
\end{equation}
and the New Newtonian potential is given by 
$\Phi_N \equiv (1/2)h_{00}=\psi-(1/2) \bar{\phi}/\langle\phi\rangle$.
Above equation together  with
\begin{equation}\label{eq:phibar}
  \nabla^2 \bar{\phi} - \lambda^{-2} \bar{\phi} = - 8\pi \alpha\rho \; ,
\end{equation}
form a Poisson-Helmholtz equation and gives
\begin{displaymath}
\Phi_N =\psi-\frac{1}{2} \frac{G_N}{1+\alpha} \bar{\phi}
\end{displaymath}
which represents
the Newtonian limit of the STT with arbitrary potential $V(\phi)$ and function
$\omega(\phi)$ that where Taylor expanded around $\langle\phi\rangle$.
The resulting equations are then distinguished by the constants
$G_N$, $\alpha$, and $\lambda=h_P/m_{SF}c$. Here $h_P$ is Planck's constant.

The next step is to find solutions for this new Newtonian potential given 
a density profile, that is, to find the so--called potential--density pairs. 
General solutions to Eqs. (\ref{eq:Psi}) and (\ref{eq:phibar}) 
can be found in terms of the corresponding Green functions,
and the new Newtonian potential is \cite{mar2004,mar2005}
%(Rodriguez-Meza et al. 2004, 2005)
\begin{eqnarray}
\Phi_N  %\equiv \frac{1}{2} h_{00}
&=& - \frac{G_N}{1+\alpha} \int d{\bf r}_s
\frac{\rho({\bf r}_s)}{|{\bf r}-{\bf r}_s|} \nonumber \\
&& -\alpha \frac{G_N}{1+\alpha} \int d{\bf r}_s \frac{\rho({\bf r}_s)
{\rm e}^{- |{\bf r}-{\bf r}_s|/\lambda}}
{| {\bf r}-{\bf r}_s|} + \mbox{B.C.} \label{pares_eq_gralPsiN}
\end{eqnarray}
The first term of Eq. (\ref{pares_eq_gralPsiN}), is the
contribution of the usual Newtonian gravitation (without SF), 
while information about the SF is contained in the
second term, that is, arising from the influence function determined by the
modified Helmholtz Green function, where the coupling $\omega$ ($\alpha$) enters
as part of a source factor.

The potential of a single particle of mass $m$ can be easily obtained from
(\ref{pares_eq_gralPsiN}) and is given by
\begin{equation}\label{eq:pointMassPotential}
\Phi_N = -\frac{G_N}{1+\alpha}\frac{m}{r} \left(
1 + \alpha e^{-r/\lambda}
\right)
\end{equation}
For local scales, $r\ll \lambda$, deviations from the Newtonian theory are exponentially
suppressed, and for $r\gg \lambda$ the Newtonian constant diminishes (augments)
to $G_N/(1+\alpha)$ for positive (negative) $\alpha$. This means that equation
(\ref{eq:pointMassPotential}) fulfills all local tests of the Newtonian dynamics, and it is
only constrained by experiments or tests on scales larger than --or of the order of--
$\lambda$, which in our case is of the order of galactic scales. 
In contrast, the potential in the form of equation (\ref{eq:YukawaPot01}) with the gravitational
constant defined as usual does not fulfills the local tests of the Newtonian dynamics
\citep{Fischback1999}.

It is appropriate to give some additional details on the Newtonian limit for
the Einstein equations without scalar fields (see \cite{Peebles1980}). 
We are considering a small region
compared to the Hubble length $cH^{-1}$ but large compared to the Schwarzschild
radii of any collapsed object. In this small region the metric tensor was written as
$g_{\mu\nu} = \eta_{\mu\nu} + h_{\mu\nu}$ where $ h_{\mu\nu}$ is small as compared to
the Minkowski metric $\eta_{\mu\nu}$. In this region Einstein's field equations are
simple because the standard weak field linear approximation applies. One finds,
\begin{eqnarray}
R_{00} &=&-\frac{1}{2} \eta^{\mu\nu} (h_{\mu\nu,00} -h_{\mu 0, \nu 0} 
-h_{\nu0,\mu 0} + h_{00,\mu\nu} ) 
\nonumber \\
&=& \nabla_r^2 \Phi
\\
g_{00} &=& c^2 + 2 \Phi \label{eq:g00}
\end{eqnarray}
Then, the zero-zero component of the field equations for an ideal fluid with density $\rho$,
pressure $p$, and velocity $v \ll c$ becomes
\begin{equation}\label{eq:Poisson_Lambda}
\nabla_r^2 \Phi = 4\pi G_N \left ( 
\rho + 3 \frac{p}{c^2} 
\right )- \Lambda
\end{equation}
For completeness the cosmological constant has been added. The geodesic equations,
in the limit $v\ll c$, $h\ll 1$, are
\begin{equation}\label{eq:Geodesic}
\frac{d^2 r^i}{dt^2} = -\Phi_{,i}
\end{equation}
Equations (\ref{eq:Poisson_Lambda}) and (\ref{eq:Geodesic}) are the standard equations of Newtonian mechanics, except
that if there is an appreciable radiation background, one must take into account the 
active gravitational mass associated with the pressure, and of course if $\Lambda \neq 0$,
there is the cosmic force $\Lambda \textbf{r}/3$ between particles at separation $\textbf{r}$.

Equations (\ref{eq:Poisson_Lambda}) and (\ref{eq:Geodesic})  apply to any observer outside
a singularity, though depending on the situation, the region within which these equations
apply need not contain much matter. The region can be extended by giving the observer
an acceleration $g_i$ to bring the observer to rest relative to distant matter, which adds the
term $g_i r^i$ to
$\Phi$, and then by patching together the results from neighboring observers. This works
(the acceleration and potentials can be added) as long as relative velocities of observers and
observed matter are $\ll c$ and $\Phi \ll c^2$ (equation (\ref{eq:g00})). For a region of size
$R$ containing mass $M\sim \rho R^3$ with density $\rho$ roughly uniform, this second
condition is
\begin{equation}\label{eq:RCondtion}
G_N\rho R^2 \ll c^2
\end{equation}
In the Friedmann-Lema\^itre models Hubble's constant is
\begin{equation}
H \sim (G_N\rho)^{1/2}
\end{equation}
If one assumes $\Lambda$ is negligible and the density parameter $\Omega\sim 1$, so 
equation (\ref{eq:RCondtion}) indicates
\begin{equation}
R\ll cH^{-1} \sim 3000 \text{ Mpc} \sim 10^{28} \text{ cm}
\end{equation}
That is, the region must be small compared to the Hubble length. Since the expansion
velocity is $v\approx Hr$, this condition also says $v\ll c$.

The Newtonian approximation can fail at much smaller $R$ if the region includes a compact
object like a neutron star or black hole, but one can deal with this by noting that at 
distances large compared to the Schwarzschild radius the object acts like an 
ordinary Newtonian point mass. It is speculated that in nuclei of galaxies there might be
black holes as massive as $10^9$ M$_\odot$, Schwarzschild radius $\sim 10^{14}$ cm.
If this is an upper limit, Newtonian mechanics is a good approximation over a substantial
range of scales, $10^{14} \text{ cm} \ll r \ll 10^{28}  \text{ cm}$.

%%%%%%%%%%%% NEWTONIAN LIMIT OF STT %%%%%%%%%%%%%
\subsection{Multipole expansion of the Poisson-Helmholtz equations}\label{sec:Multipole}

The Poisson's Green function can be expanded in terms of the spherical 
harmonics, $Y_{ln}(\theta,\varphi)$,
\begin{displaymath}
\frac{1}{|{\bf r}-{\bf r}_s|}=4\pi \sum_{l=0}^\infty \sum_{n=-l}^l \frac{1}{2l+1}
\frac{r_<^l}{r_>^{l+1}} Y_{ln}^*(\theta',\varphi') Y_{ln}(\theta,\varphi), 
\end{displaymath}
where $r_<$ is the smaller of $|{\bf r}|$ and $|{\bf r}_s|$, and
$r_>$ is the larger of $|{\bf r}|$ and $|{\bf r}_s|$ and it allows us that
the standard gravitational potential due to a distribution of mass $\rho({\bf r})$,
without considering the boundary
condition, can be written as   \citep{Jackson1975}
\begin{displaymath}
\psi({\bf r}) = \psi^{(i)}+\psi^{(e)}
\end{displaymath}
where $\psi^{(i)}$ ($\psi^{(e)}$) are the internal (external) multipole expansion of 
$\psi$,
\begin{eqnarray}
\psi^{(i)} &=& - \sum_{l=0}^\infty \sum_{n=-l}^l \frac{\sqrt{4\pi}}{2l+1} q_{ln}^{(i)} 
Y_{ln}(\theta,\varphi) r^{l} \, ,  \nonumber \\
\psi^{(e)} &=& - \sum_{l=0}^\infty \sum_{n=-l}^l \frac{\sqrt{4\pi}}{2l+1} q_{ln}^{(e)} 
\frac{Y_{ln}(\theta,\varphi)}{r^{l+1}} \, , \nonumber
\end{eqnarray}

Here, the 
coefficients of the expansions $\psi^{(i)}$ and $\psi^{(e)}$, known as internal and external 
multipoles, respectively, are given by
\begin{eqnarray}\label{imultipoles_psi_eq}
q_{ln}^{(i)} &=&  \sqrt{4\pi} \int_{V(r\le r')} d{\bf r}' \frac{1}{r'^{l+1}} Y_{ln}^*(\theta',\varphi') \rho({\bf r}')  \, , \nonumber \\
q_{ln}^{(e)} &=& \sqrt{4\pi}  \int_{V(r>r')} d{\bf r}' Y_{ln}^*(\theta',\varphi') r'^l \rho({\bf r}')  \, .
\nonumber
\end{eqnarray}

%%%%%%%%%%%%%%%%%%%%%%%%%%%%%%%%%%%%%%%%%%%%%
The integrals are done in a region $V$ where $r\le r'$ for the internal multipoles and in a region
$V$ where $r>r'$ for the external multipoles.
They have the property
\begin{eqnarray}
\begin{array}{l}
q_{l(-n)}^{(i)}=(-1)^n (q_{ln}^{(i)})^* \\[.1in]
q_{l(-n)}^{(e)}=(-1)^n (q_{ln}^{(e)})^*  
\end{array}
\end{eqnarray}

We may write expansions above in cartesian coordinates up to quadrupoles. For the internal
multipole expansion we have
\begin{equation}\label{cartesian_ipsi_eq}
\psi^{(i)} = -M^{(i)} - \mathbf{r}\cdot \mathbf{p}^{(i)}-\frac{1}{2} 
\mathbf{r}\cdot \mathbf{Q}^{(i)}\cdot\mathbf{r} \, ,
\end{equation}
and its force is
\begin{equation}\label{cartesian_iFpsi_eq}
\mathbf{F}_\psi^{(i)} = \mathbf{p}^{(i)}+ 
\mathbf{Q}^{(i)}\cdot\mathbf{r} \, ,
\end{equation}
where
\begin{equation}
M^{(i)}\equiv \int_{V(r\le r')} d{\bf r}' \frac{1}{r'}  \rho({\bf r}')  \, ,
\end{equation}
\begin{equation}
p_i^{(i)}\equiv \int_{V(r\le r')} d{\bf r}' \, x'_i \frac{1}{r'^3}  \rho({\bf r}')  \, ,
\end{equation}
\begin{equation}
Q_{ij}^{(i)}\equiv \int_{V(r\le r')} d{\bf r}' \, (3x'_i x'_j-r'^2 \delta_{ij} ) \frac{1}{r'^5}  \rho({\bf r}')  \, .
\end{equation}

For the external multipoles we have
\begin{equation}\label{cartesian_epsi_eq}
\psi^{(e)} = -\frac{M^{(e)}}{r} - \frac{\mathbf{r}\cdot \mathbf{p}^{(e)}}{r^3}
-\frac{1}{2} \frac{\mathbf{r}\cdot \mathbf{Q}^{(e)}\cdot\mathbf{r}}{r^5} \, ,
\end{equation}
and its force is
\begin{eqnarray}\label{cartesian_eFpsi_eq}
\mathbf{F}_\psi^{(e)} &=& -\frac{M^{(e)}}{r^3}\mathbf{r} + \frac{\mathbf{p}^{(e)}}{r^3}
-3\frac{\mathbf{p}^{(e)}\cdot\mathbf{r}}{r^5}\mathbf{r} \nonumber \\
&& + \frac{\mathbf{Q}^{(e)}\cdot\mathbf{r}}{r^5}
-\frac{5}{2} \frac{\mathbf{r}\cdot\mathbf{Q}^{(e)}\cdot\mathbf{r}}{r^7}\mathbf{r}
 \, ,
\end{eqnarray}
where
\begin{equation}
M^{(e)}\equiv \int_{V(r> r')} d{\bf r}'  \rho({\bf r}')  \, ,
\end{equation}
\begin{equation}
p_i^{(e)}\equiv \int_{V(r> r')} d{\bf r}' \, x'_i   \rho({\bf r}')  \, ,
\end{equation}
\begin{equation}
Q_{ij}^{(e)}\equiv \int_{V(r> r')} d{\bf r}' \, (3x'_i x'_j-r'^2 \delta_{ij} )  \rho({\bf r}')  \, .
\end{equation}

The external multipoles have the usual meaning, i.e., $M^{(e)}$ is the mass,
$\mathbf{p}^{(e)}$ is the dipole moment, and $\mathbf{Q}^{(e)}$ is the traceless quadrupole
tensor, of the volume $V(r>r')$.
We may atach to the internal multipoles similar meaning, i.e., $M^{(i)}$ is the internal ``mass'',
$\mathbf{p}^{(i)}$ is the internal ``dipole'' moment, and $\mathbf{Q}^{(i)}$ is the traceless internal
``quadrupole'' tensor, of the volume $V(r\le r')$.
%%%%%%%%%%%%%%%%%%%%%%%%%%%%%%%%%%%%%%%%%%%%%

In the case of the scalar field, with the expansion
\begin{eqnarray}
\frac{\exp(-m |{\bf r}-{\bf r}_s|)}{|{\bf r}-{\bf r}_s|}&=&
4\pi m \sum_{l=0}^\infty \sum_{n=-l}^l 
i_l(m r_<) k_l(m r_>) 
\nonumber \\ && \times
Y_{ln}^*(\theta',\varphi') Y_{ln}(\theta,\varphi) \; ,  \nonumber
\end{eqnarray}
the contribution of the scalar field to the Newtonian gravitational potential 
can be written as 
\begin{displaymath}
\bar{\phi}({\bf r}) = \bar{\phi}^{(i)}+\bar{\phi}^{(e)}
\end{displaymath}
where, for simplicity of notation, we are using $m=m_{SF}=h_P/(c \lambda)$ and
\begin{eqnarray}
\frac{1}{2\alpha} \bar{\phi}^{(i)} &=& \sqrt{4\pi}
\sum_{l=0}^\infty \sum_{n=-l}^l  \frac{i_l(m r)}{(m r)^{l}} \,  \bar{q}_{ln}^{(i)} 
r^{l}Y_{ln}(\theta,\varphi) \, , \nonumber \\
\frac{1}{2\alpha} \bar{\phi}^{(e)} &=& \sqrt{4\pi}
\sum_{l=0}^\infty \sum_{n=-l}^l (m r)^{l+1} k_l(m r) \,  \bar{q}_{ln}^{(e)} 
\frac{Y_{ln}(\theta,\varphi)}{r^{l+1}} \, , \nonumber
\end{eqnarray}
$i_l(x)$ and $k_l(x)$ are the modified spherical Bessel functions.

We 
have defined the multipoles for the scalar field as
\begin{eqnarray}\label{imultipoles_phi_eq}
\bar{q}_{ln}^{(i)} &=& \sqrt{4\pi} \int_{V(r\le r')} d{\bf r}' \, 
\frac{Y_{ln}^*(\theta',\varphi')}{r'^{l+1}}\, (m r')^{l+1} k_l(m r') \, \rho({\bf r}') , \nonumber \\
\bar{q}_{ln}^{(e)} &=& \sqrt{4\pi} \int_{V(r>r')} d{\bf r}' \, Y_{ln}^*(\theta',\varphi')\, 
\frac{i_l(m r')}{(m r')^l} r'^l\, \rho({\bf r}') \, . \nonumber
\end{eqnarray}
%}

%%%%%%%%%%%%%%%%%%%%%%%%%%%%%%%%%%%%%%%%%%%%%%
They, also, have the property
\begin{eqnarray}
\begin{array}{l}
\bar{q}_{l(-n)}^{(i)}=(-1)^n (\bar{q}_{ln}^{(i)})^* \\[.1in]
\bar{q}_{l(-n)}^{(e)}=(-1)^n (\bar{q}_{ln}^{(e)})^*  
\end{array}
\end{eqnarray}

The above expansions of SF contribution to the Newtonian potential can be written
in cartesian coordinates. The internal multipole expansion of the SF contribution, 
up to quadrupoles is
\begin{eqnarray}
\frac{1}{2\alpha} \bar{\phi}^{(i)} &=& i_0(mr) \bar{M}^{(i)} 
+ 3\frac{i_1(m r)}{m r} \mathbf{r}\cdot \bar{\mathbf{p}}^{(i)} 
\nonumber 
\\
&&+5\frac{1}{2} \frac{i_2(m r)}{(m r)^2} \mathbf{r}\cdot \bar{\mathbf{Q}}^{(i)}\cdot\mathbf{r}
\label{cartesian_iphi_eq}
\end{eqnarray}
and its force is
\begin{eqnarray}
\frac{1}{2\alpha}\mathbf{F}_\phi^{(i)} &=&
-m^2 \frac{i_1(mr)}{mr} \bar{M}^{(i)} \mathbf{r} 
-3 \frac{i_1(mr)}{mr} \bar{\mathbf{p}}^{(i)}  
%\nonumber 
\\
&&-3 m^2 \frac{i_2(mr)}{(mr)^2} (\bar{\mathbf{p}}^{(i)}\cdot\mathbf{r}) \mathbf{r}
-5 \frac{i_2(mr)}{(mr)^2} \bar{\mathbf{Q}}^{(i)}\cdot\mathbf{r}   \label{cartesian_iFphi_eq} 
\nonumber
\\
&&+5\frac{1}{2r^2} [ 5\frac{i_2(mr)}{(mr)^2} - \frac{i_1(mr)}{mr} ]
(\mathbf{r}\cdot\bar{\mathbf{Q}}^{(i)}\cdot\mathbf{r}) \mathbf{r}
\, ,	\nonumber
\end{eqnarray}
where
\begin{equation}
\bar{M}^{(i)}\equiv \int_{V(r\le r')} d{\bf r}'  (mr') k_0(mr')
\frac{1}{r'}  \rho({\bf r}')  \, ,
\end{equation}
\begin{equation}
\bar{p}_i^{(i)}\equiv \int_{V(r\le r')} d{\bf r}' \, (mr')^2 k_1(mr')
x'_i \frac{1}{r'^3}  \rho({\bf r}')  \, ,
\end{equation}
\begin{equation}
\bar{Q}_{ij}^{(i)}\equiv \int_{V(r\le r')} d{\bf r}' \, (mr')^3 k_2(mr')
(3x'_i x'_j-r'^2 \delta_{ij} ) \frac{1}{r'^5}  \rho({\bf r}')  \, .
\end{equation}

In the exterior region the SF multipole contribution to the potential is 
\begin{eqnarray}
\frac{1}{2\alpha} \bar{\phi}^{(e)} &=& mr\, k_0(mr) \frac{\bar{M}^{(e)}}{r}
+ 3 (mr)^2 k_1(m r) \frac{\mathbf{r}\cdot \bar{\mathbf{p}}^{(e)}}{r^3} 
\nonumber \\
&& + 5\frac{1}{2} (mr)^3 k_2(m r) \frac{\mathbf{r}\cdot \bar{\mathbf{Q}}^{(e)}\cdot\mathbf{r}}{r^5}
\label{cartesian_ephi_eq}
\end{eqnarray}
and its force is
\begin{eqnarray}
\frac{1}{2\alpha}\mathbf{F}_\phi^{(e)} &=&
(mr)^2 k_1(mr) \frac{\bar{M}^{(e)}}{r^3} \mathbf{r} 
-3 (mr)^2 k_1(mr) \frac{\bar{\mathbf{p}}^{(e)}}{r^3}  \nonumber \\
&&+3 \left[ (mr) k_0(mr)
\right. 
\nonumber \\
&&\left.
+(2+mr)(mr)^2 k_1(mr) \right] 
\frac{(\bar{\mathbf{p}}^{(e)}\cdot\mathbf{r})}{r^5} \mathbf{r} \nonumber \\
&&
-5 (mr)^3 k_2(mr) \frac{\bar{\mathbf{Q}}^{(e)}\cdot\mathbf{r}}{r^5} 
\label{cartesian_eFphi_eq} \\
&&+\frac{5}{2} \left[ 3(mr) k_0(mr) + 3(mr)^2 k_1(mr) \right. \nonumber \\
&&+\left. (3+mr)(mr)^3 k_2(mr) \right]
\frac{(\mathbf{r}\cdot\bar{\mathbf{Q}}^{(e)}\cdot\mathbf{r})}{r^7} \mathbf{r}
\, ,	\nonumber
\end{eqnarray}
where
\begin{equation}
\bar{M}^{(e)}\equiv \int_{V(r> r')} d{\bf r}'  i_0(mr') \rho({\bf r}')  \, ,
\end{equation}
\begin{equation}
\bar{p}_i^{(e)}\equiv \int_{V(r> r')} d{\bf r}' \,  \frac{i_1(mr')}{mr'}
x'_i  \rho({\bf r}')  \, ,
\end{equation}
\begin{equation}
\bar{Q}_{ij}^{(e)}\equiv \int_{V(r> r')} d{\bf r}' \, \frac{i_2(mr')}{(mr')^2 }
(3x'_i x'_j-r'^2 \delta_{ij} ) \rho({\bf r}')  \, .
\end{equation}

In the limit when $m\rightarrow 0$ we recover the standard Newtonian potential and force
expressions.

Up to here the formulation is general, i.e., mass distribution may have any symmetry or none at all.
In order to take advantage of the symmetry of the spherical harmonics, the mass distribution must 
be spherically symmetric.
%%%%%%%%%%%%%%%%%%%%%%%%%%%%%%%%%%%%%%%%%%%%%%

%%%%%%%%%%%% TEORIA ESCALAR TENSORIAL - COSMOS %%%%%%%%%%%%%

\section{Cosmological evolution equations using a static STT}\label{sec:Cosmos}

To simulate cosmological systems,  the expansion of the universe has to be
taken into account.
Also, to determine the nature of the cosmological model we need to determine
 the composition of the
universe, i. e., we need to give the values of $\Omega_i\equiv \rho_i/\rho_c$, with
$\rho_c=3H^2/8\pi G_N$, 
for each component $i$, 
taking into account
in this way all forms of energy densities that exist at present.
If a particular kind of energy density is described by an equation of state of the form
$p=w \rho$, where $p$ is the pressure and $w$ is a constant, then the equation for energy
conservation in an expanding background, $d(\rho a^3)=-pd(a^3)$, can be integrated to
give $\rho \propto a^{-3(1+w)}$.

Then, the Friedmann equation for the expansion factor $a(t)$
is written as
\begin{equation}
\frac{\dot{a}^2}{a^2} = H_0^2 \sum_i \Omega_i \left(\frac{a_0}{a}\right)^{3(1+w_i)} - \frac{k}{a^2}
\end{equation}
where $w_i$ characterizes equation of state of species $i$. 

The most familiar forms of energy
densities are those due to pressureless matter with $w_i=0$ (that is, nonrelativistic matter
with rest-mass-energy density $\rho c^2$ dominating over the kinetic-energy density
$\rho v^2/2$) and radiation with $w_i=1/3$.  The density parameter contributed today
by visible, nonrelativistic, baryonic matter in the universe is $\Omega_B \approx (0.04-0.05)$
and the density parameter that is due to radiation is $\Omega_R \approx 2\times 10^{-5}$.

In this work we will consider a model with only two energy density contribution.
One which is a pressureless and 
nonbaryonic dark matter  with $\Omega_{DM} \approx 0.3$ that does not couple with radiation.
Although in the numerical simulations we may include in $\Omega_{DM}$ the baryonic matter.
Other, that will be a cosmological constant contribution $\Omega_\Lambda \approx 0.7$
with and equation of state $p =-\rho$. The above equation for $a(t)$ becomes
\begin{equation}
\frac{\dot{a}^2}{a^2} = H_0^2 
\left[ 
\Omega_{DM} \left(\frac{a_0}{a}\right)^{3} +  \Omega_\Lambda
\right]
- \frac{k}{a^2}
\end{equation}

The above discussion gives us the standard cosmological model with cosmological constant,
i. e., $\Lambda$CDM model. 

In the framework of a
scalar-tensor theory the cosmology is given as follows.
If we use the Friedmann metric \cite{Faraoni2004}
\begin{equation}
ds^2=-dt^2+a^2(t)
\left[
\frac{dr^2}{1-\kappa r^2} + r^2 (d\theta^2 + \sin^2\theta d\varphi^2)
\right]
\end{equation}
in the time-time component of the Einstein field equations (Hamiltonian constraint)
gives
\begin{equation}
H^2=\frac{8\pi}{3 \phi} \rho - H \frac{\dot{\phi}}{\phi}
+\frac{\omega(\phi)}{6}\left( \frac{\dot{\phi}}{\phi} \right)^2
-\frac{\kappa}{a^2}+\frac{V}{6\phi}
\end{equation}
while the equation for the scalar field is
\begin{eqnarray}
\ddot{\phi} 
+ \left( 3H+\frac{\dot{\omega}}{2\omega+3} \right) \dot{\phi} = 
\nonumber \\
\frac{1}{2\omega+3} 
\left[
8\pi \left( \rho -3 P \right) -\phi \frac{dV}{d\phi}+2V
\right]
\end{eqnarray}
and the equation of the fluid is
\begin{equation}
\dot{\rho}+3H  \left( \rho + P \right) = 0
\end{equation}

The cosmological evolution of the initial perturbed fields should be 
computed using the above equations for the expansion factor.
However,
here,
we will  employ a cosmological model with a static SF which is consistent with the 
Newtonian limit given above.
Thus, the scale factor, $a(t)$,  is given by the following Friedman model,
\begin{equation} \label{new_friedman}
a^3 H^2= H_{0}^{2} \left[\frac{\Omega_{DM0} +  \Omega_{\Lambda 0} \, a^3}{1+\alpha} 
+  \left(1-\frac{\Omega_{DM 0}+\Omega_{\Lambda 0}}{1+\alpha} \right) \, a  \right]
\end{equation}
where $H=\dot{a}/a$,  $\Omega_{DM 0}$ and $\Omega_{\Lambda 0}$ 
are the matter and energy density evaluated at present, respectively.   
The denominator $(1+\alpha)$ appears in Eq. (\ref{new_friedman}) due to the fact that
we have defined $\Omega_i\equiv \rho/\rho_c$ and $\rho_c=3 H^2/8\pi G_N$. Then,
whenever appears the gravitational constant $G_N$, we replaces it by $G_N/(1+\alpha)$.

We notice that the source of the cosmic evolution is deviated by the term 
$1+\alpha$ when compared to the standard Friedmann-Lema\^itre 
model. Therefore, it is convenient to define a new density parameter by 
$\Omega_i^{(\alpha)} \equiv \Omega_i/(1+\alpha)$. This new density parameter is such that 
$\Omega_{DM}^{(\alpha)} + \Omega_\Lambda^{(\alpha)} =1$, 
which implies a flat universe, and this shall be assumed 
in the following computations, where we consider 
$(\Omega_m^{(\alpha)}, \Omega_\Lambda^{(\alpha)}) = (0.3, 0.7) $.  

\begin{figure}
\begin{center}
\includegraphics[width=3.8in]{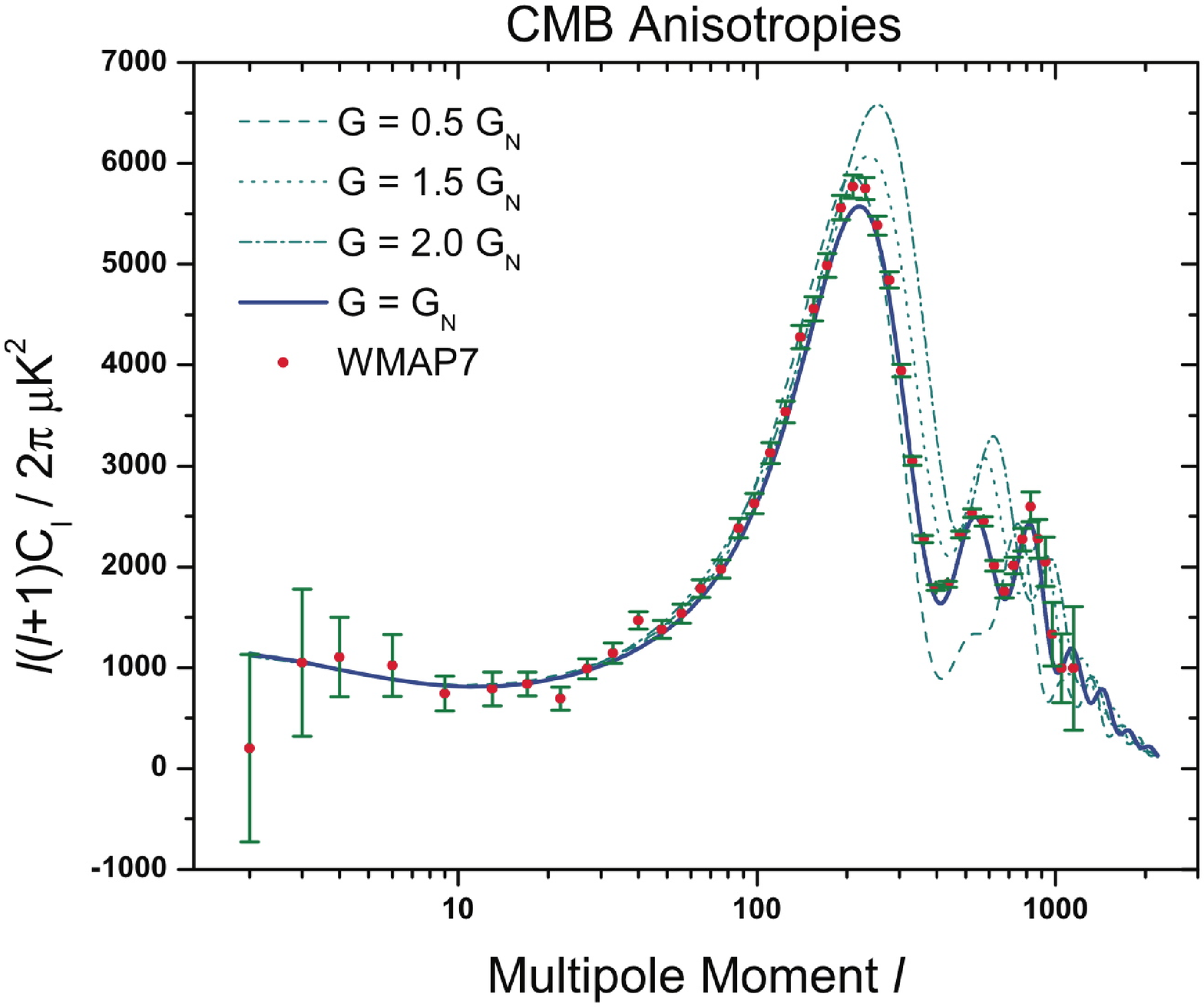}
\end{center}
\caption{Cosmic Microwave Background anisotropies. With small dash symbol are the 
WMAP 7 years experimental observations \cite{Spergel2003}. We have plotted several cases in which 
the Newton constant has been modified. The standard case is shown with a small dash
line. With a long dash line is shown an effective gravitational constant, 
$G_{eff}=0.5 G_N$. With a dotted line
is shown the case with $G_{eff}=1.5 G_N$, and with a dash--dotted line is shown
the case with $G_{eff}=2 G_N$.}\label{fig:CMB_moG}
\end{figure}

For positive values 
of $\alpha$, a flat cosmological model demands to have a factor $(1+\alpha)$ more energy 
content ($\Omega_m$ and $ \Omega_\Lambda$) than in standard cosmology. 
On the other hand, for negative values of  
$\alpha$ one needs a factor $(1+\alpha)$  less $\Omega_m$ 
and $ \Omega_\Lambda$ to have a flat universe.  
To be consistent 
with the CMB  spectrum and structure formation numerical 
experiments, cosmological constraints must be applied on $\alpha$ in order for it to 
be within the range $(-1,1)$ 
\cite{Nagata2002,Nagata2003,Shirata2005,Umezu2005}.
In figure \ref{fig:CMB_moG} we show the effect of different values of the gravitational
constant $G_N$ on the anisotripies of the CMB.

%%%%%%%%%%%%%%% NBODY %%%%%%%%%%%%%%%%

\section{Vlasov-Poisson-Helmholtz equations and the $N$-body method}\label{sec:Nbody}

The Vlasov-Poisson equation in an expanding universe describes the evolution of the
six-dimensional, one-particle distribution function, $f(\mathbf{x},\mathbf{p})$.
The Vlasov equation is given by,
\begin{equation}\label{Vlasov_eq}
\frac{\partial f}{\partial t} + \frac{\mathbf{p}}{m a^2} \cdot \frac{\partial f}{\partial \mathbf{x}} 
-  m \nabla \Phi_N (\mathbf{x}) \cdot \frac{\partial f}{\partial \mathbf{p}} = 0 
\end{equation}
where $\mathbf{x}$ is the comoving coordinate, 
$\mathbf{p}=m a^2 \dot{\mathbf{x}}$, $m$ is the particle mass, and $\Phi_N$ is
the self-consistent gravitational potential given by the
Poisson equation,
\begin{equation}\label{Poisson_eq}
\nabla^2 \Phi_N(\mathbf{x}) = 4 \pi G_N \, a^2 
\left[
\rho(\mathbf{x}) - \rho_b(t) %]
\right]
\end{equation}
where $\rho_b$ is the background mass density. 
The Vlasov-Poisson system,
Eqs. (\ref{Vlasov_eq}) and (\ref{Poisson_eq}) form the Vlasov-Poisson equation,
constitutes a collisionless, mean-field approximation to the evolution of the full
$N$-body distribution.

An $N$-body code attempts to solve the Vlasov-Poisson system of equations
by representing the one-particle
distribution function as
\begin{equation}\label{eq:DiscreteDistribution}
f(\mathbf{x},\mathbf{p}) = \sum_{i=1}^N \delta(\mathbf{x}-\mathbf{x}_i)\, 
\delta(\mathbf{p}-\mathbf{p}_i)
\end{equation}
Substitution of (\ref{eq:DiscreteDistribution}) in the Vlasov-Poisson system of 
equations yields the exact Newton's equations for a system of $N$ gravitating
particles 
(See \cite{Bertschinger1998} for details),
\begin{equation} \label{eq_motion_newton}
\ddot{\mathbf{x}}_i + 2\, H \, \mathbf{x}_i 
= 
-\frac{1}{a^3} 
G_N
\sum_{j\ne i} \frac{m_j (\mathbf{x}_i-\mathbf{x}_j)}
{|\mathbf{x}_i-\mathbf{x}_j|^3} 
\end{equation}
where the sum includes all  
periodic images of particle $j$ and numerically is done using the Ewald
method, see \cite{Hernquist1991}. 
It is important to keep in mind, however, that we are not really interested in solving
the exact $N$-body problem for a finite number of particles, $N$.
The actual problem of interest is the exact $N$-body problem in the fluid limit, i. e.,
as $N\rightarrow \infty$. For this reason, one important aspect of the 
numerical fidelity of $N$-body codes lies in controlling errors due to the discrete
nature of the particle representation of $f(\textbf{r},\textbf{p})$.

In the Newtonian limit of STT of gravity to describe the evolution of the
six-dimensional, one-particle distribution function, $f(\mathbf{x},\mathbf{p},t)$
we need to solve
the Vlasov-Poisson-Helmholtz equation in an expanding universe
\citep{mar2007,mar2008,mar2009,mar2009b,mar2010}.
The Vlasov equation is given by
\begin{equation}\label{Vlasov_eq}
\frac{\partial f}{\partial t} + \frac{\mathbf{p}}{m a^2} \cdot \frac{\partial f}{\partial \mathbf{x}} 
-  m \nabla \Phi_N (\mathbf{x}) \cdot \frac{\partial f}{\partial \mathbf{p}} = 0 
\end{equation}
where 
$\Phi_N = (1/2)h_{00}=\psi-\frac{1}{2} \frac{G_N}{1+\alpha} \bar{\phi}$ 
with $\psi$ satisfying Poisson equation
\begin{equation}
\nabla^2 \psi(\mathbf{x}) = 4 \pi G_N \, 
a^2 
\left[
\rho(\mathbf{x}) - \rho_b(t)   %]
\right]
\end{equation}
and $\bar{\phi}$ satisfying the Helmholtz equation
\begin{equation}
  \nabla^2 \bar{\phi} - \lambda^{-2} \bar{\phi} = - 8\pi \alpha
  a^2 
\left[
\rho(\mathbf{x}) - \rho_b(t)  %]
\right]
 \; ,
\end{equation}
Above equations  form the Vlasov-Poisson-Helmholtz system of equations,
constitutes a collisionless, mean-field approximation to the evolution of the full
$N$-body distribution in the framework of the Newtonian limit of a scalar-tensor theory.

Using the representation of the one-particle
distribution function (\ref{eq:DiscreteDistribution})
the Newtonian motion equation  for a particle $i$, is written as \citep{mar2008}
\begin{eqnarray} \label{eq_motion}
\ddot{\mathbf{x}}_i + 2\, H \, \mathbf{x}_i &=& 
-\frac{1}{a^3} \frac{G_N}{1+\alpha} \sum_{j\ne i} \frac{m_j (\mathbf{x}_i-\mathbf{x}_j)}
{|\mathbf{x}_i-\mathbf{x}_j|^3} 
\nonumber \\ &&
\times F_{SF}(|\mathbf{x}_i-\mathbf{x}_j|,\alpha,\lambda)
\end{eqnarray}
where the sum includes all  
periodic images of particle $j$,  and $F_{SF}(r,\alpha,\lambda)$ is
\begin{equation}\label{eq:FSFFactor}
F_{SF}(r,\alpha,\lambda) = 1+\alpha \, \left( 1+\frac{r}{\lambda} \right)\, e^{-r/\lambda}
\end{equation}
which,  for small distances compared to $\lambda$,  is 
$F_{SF}(r<\lambda,\alpha,\lambda) \approx 1+\alpha \, \left( 1+\frac{r}{\lambda} \right)$ and, 
for long 
distances, is  $F_{SF}(r>\lambda,\alpha,\lambda) \approx 1$, as in Newtonian physics.

\begin{figure}[h]
\begin{center}
\includegraphics[width=3in]{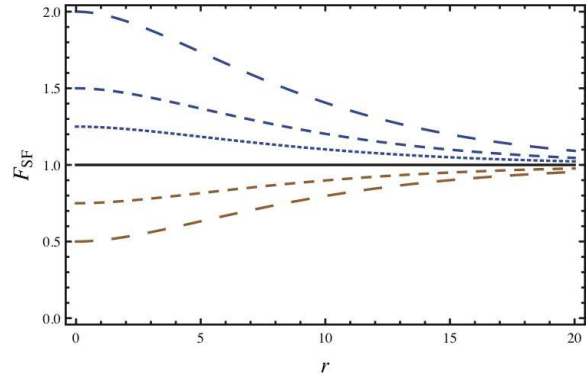}
\end{center}
\caption{Function $F_{SF}(r,\alpha,\lambda)$ for $\lambda=5$ Mpc$/h$ and several values
of parameter $\alpha$.
$\alpha=0$, i. e., the standard Newtonian case is the horizontal line. The long dash line above the horizontal standard line is for 
$\alpha=1$. 
 The medium dash line is for $\alpha=1/2$ and the tiny dash line is for $\alpha=1/4$.
For negative values of $\alpha$ the lines are below the standard horizontal case. 
Long dash line is for $\alpha=-1/2$ and the medium dash line  is for $\alpha=-1/4$.
}\label{fig:FSF}
\end{figure}

The function $F_{SF}(r,\alpha,\lambda)$ is shown in Fig. \ref{fig:FSF} 
for several values of parameter
$\alpha$. The horizontal line at $F_{SF}=1$ is for the standard Newtonian case 
($\alpha=0$). The long dash line above the horizontal standard line is for $\alpha=1$, the medium dash line is for $\alpha=1/2$ and the tiny dash line is for $\alpha=1/4$. For
negative values of $\alpha$ the lines are below the standard horizontal case. Long dash 
line is
for $\alpha=-1/2$ and medium dash line  is for $\alpha=-1/4$. 
We should notice that even though $\lambda=5$ Mpc$/h$, $F_{SF}$ gives an important 
contribution to the force between particles for $r>\lambda$.

%%%%%%%%%%%%%%%% RESULTADOS %%%%%%%%%%%%%%%%

\section{Results}\label{sec:Results}

%\subsection{Large scale structure formation}

In this section, we present results of cosmological simulations of a $\Lambda$CDM universe
with and without SF contribution in order to study the large scale structure formation. 

We have used the standard Zel'dovich approximation \citep{Zeldovich1970}
to provide the initial $256^3$ particles displacement off a uniform grid and to assign their initial
velocities
in a $256 \, h^{-1}$ Mpc box\cite{Heitmann2005}. 
In this approximation the comoving and the lagrangian coordinates
are related by
\begin{eqnarray}
\mathbf{x} &=& \mathbf{q} - 
\alpha_Z \sum_\mathbf{k} b_\mathbf{|k|}(t) \mathbf{S}_\mathbf{|k|}(\mathbf{q}), \qquad
\\
\mathbf{p} &=&  - \alpha_Z a^2 \sum_\mathbf{k} b_\mathbf{|k|}(t) 
\left( \frac{\dot{b}_\mathbf{|k|}(t)}{b_\mathbf{|k|}(t)} \right) 
\mathbf{S}_\mathbf{|k|}(\mathbf{q}),
\end{eqnarray}
where the displacement vector $\mathbf{S}$ is related to the velocity potential $\Phi$ and the
power spectrum of fluctuations $P(|k|)$:
\begin{eqnarray}
\mathbf{S}_\mathbf{|k|}(\mathbf{q}) &=& \nabla_q \Phi_\mathbf{|k|}(\mathbf{q}), \qquad \\
\Phi_\mathbf{|k|} &=& \sum_\mathbf{k} \left(
a_\mathbf{k} \cos(\mathbf{k}\cdot\mathbf{q}) +
b_\mathbf{k} \sin(\mathbf{k}\cdot\mathbf{q})
\right)
\end{eqnarray}
where $a$ and $b$ are gaussian random numbers with the mean zero and dispersion
$\sigma^2 = P(k)/k^4$,
\begin{equation}
a_\mathbf{k} =\frac{\sqrt{P(|k|)}}{|k|^2} G(0,1), \qquad
b_\mathbf{k} =\frac{\sqrt{P(|k|)}}{|k|^2} G(0,1)
\end{equation}
where $G(0,1)$ is a gaussian number with mean zero and dispersion 1.

The parameter $\alpha_Z$ together with the power spectrum $P(k)$, define the normalization
of the fluctuations. The initial power spectrum was generated using the fitting formula by
\cite{KlypinHoltzman1997} for the transfer function. This formula is a slight variation of the 
common BBKS fit \citep{Bardeen1986}.

Therefore, for the standard $\Lambda$CDM we have for the initial condition that
the starting redshift is $z_{in}=50$ and we choose the following cosmology:
$\Omega_{DM}=0.314$ (where $\Omega_{DM}$ includes cold dark matter and baryons),
$\Omega_{B}=0.044$, $\Omega_{\Lambda}=0.686$, $H_0=71$ km/s/Mpc, $\sigma_8=0.84$,
and $n=0.99$. 
Particle masses are in the order of $1.0\times 10^{10}$ M$_\odot$. 
The individual softening length
was 50 kpc$/h$. This choice of softening length is consistent
with the mass resolution set by the number of particles.
All these values are in concordance with measurements of cosmological
parameters
derived from the seven-year data of the 
 WMAP \cite{Komatsu2010}.
%\citep{Spergel2003}. 
%
The initial condition --called the big box case-- is in the Cosmic Data Bank web page:
(\url{http://t8web.lanl.gov/people/heitmann/test3.html}). 
See \cite{Heitmann2005} for more details.

Because the visible component is the smaller one and given our interest to
test the consequences of including a SF contribution to the evolution equations,
our model excludes gas particles, but all its mass has been added to the dark matter. 
We restrict the values of $\alpha$ to the interval $(-1,1)$ 
(see \cite{Nagata2002,Nagata2003,Shirata2005,Umezu2005})
 and  use $\lambda=1, 5, 10, 20$  Mpc$/h$, since 
these values sweep the scale lengths present in the simulations.

In all the simulations we have done, with or without the scalar field contribution, we demand that
the cosmological model be flat. In this model with a scalar field a flat universe is obtained if 
$\Omega_{m}^{(\alpha)}+\Omega_{\Lambda}^{(\alpha)}=1$, 
with $\Omega_{m}^{(\alpha)}\equiv \Omega_{m}/(1+\alpha)$ and 
$\Omega_{\Lambda}^{(\alpha)}\equiv \Omega_{\Lambda}/(1+\alpha)$.
 Then, for positive values of $\alpha$ we need a factor $(1+\alpha)$ more
energy content ($\Omega_m$ and $\Omega_\Lambda$) that in the standard cosmology. 
Whereas for negative values of $\alpha$ we need a factor of $(1+\alpha)$ less 
$\Omega_m$ and $\Omega_\Lambda$ to have a flat universe. With this recipe and for a given value of $\alpha$ 
we modified the
above $\Lambda$CDM initial condition accordingly.

\begin{figure}
\includegraphics[width=3.5in]{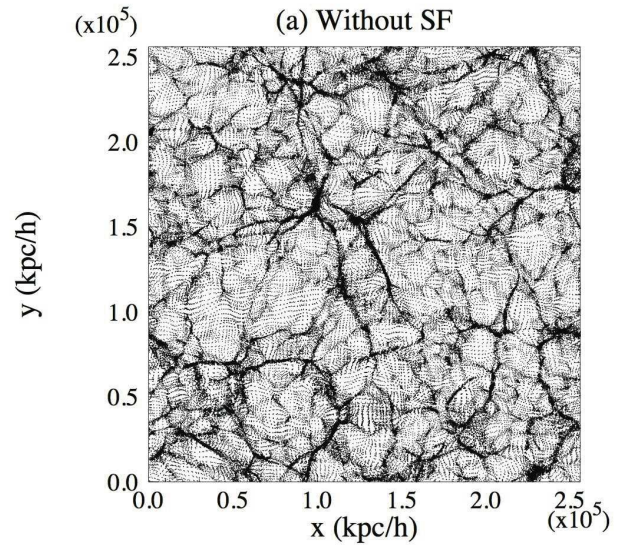}
\caption{(a) LSS formation process without the SF contribution.}
\label{fig:LSSF_a}
\end{figure}

\setcounter{figure}{2}
\begin{figure}
\includegraphics[width=3.5in]{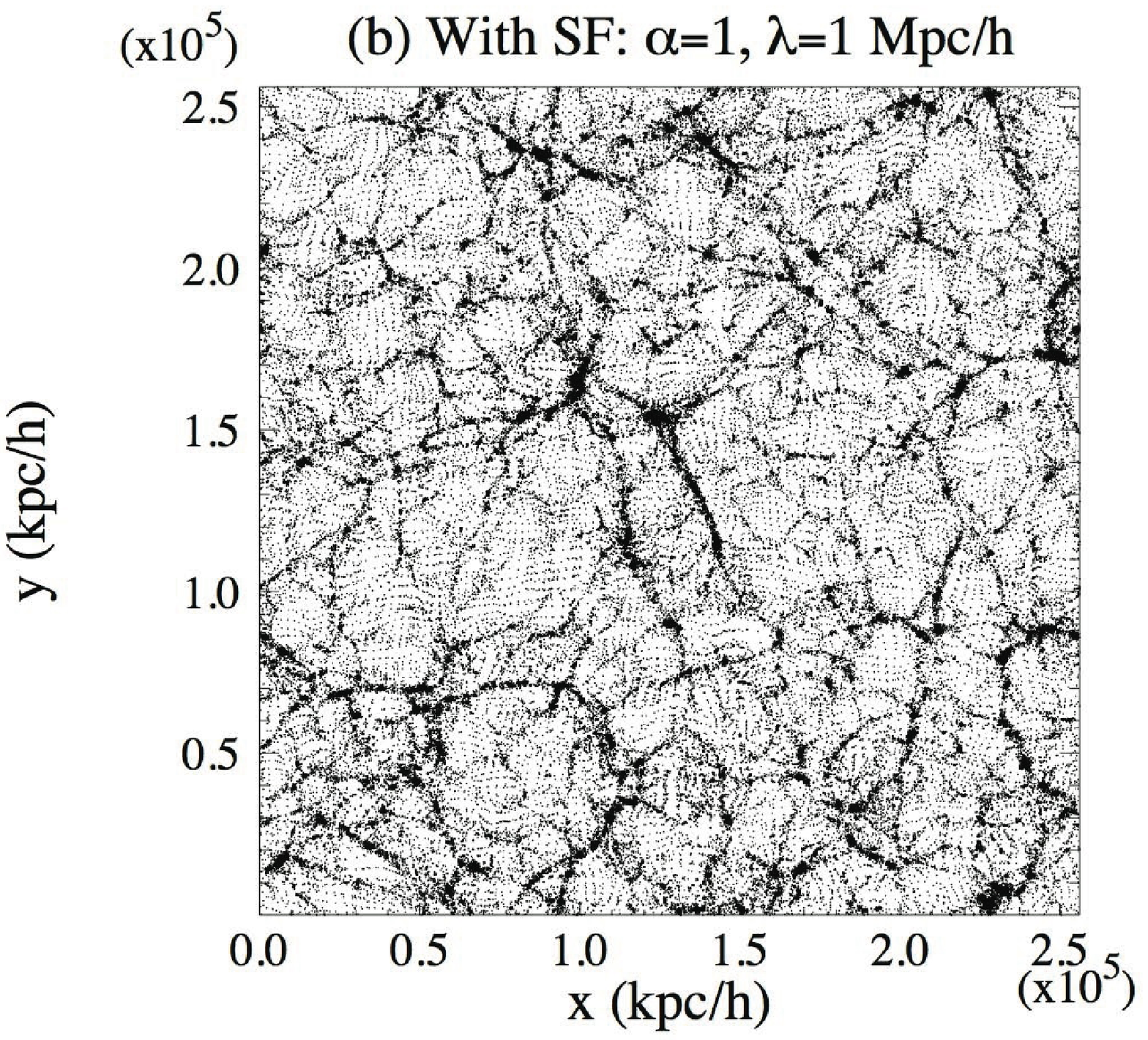}
\includegraphics[width=3.5in]{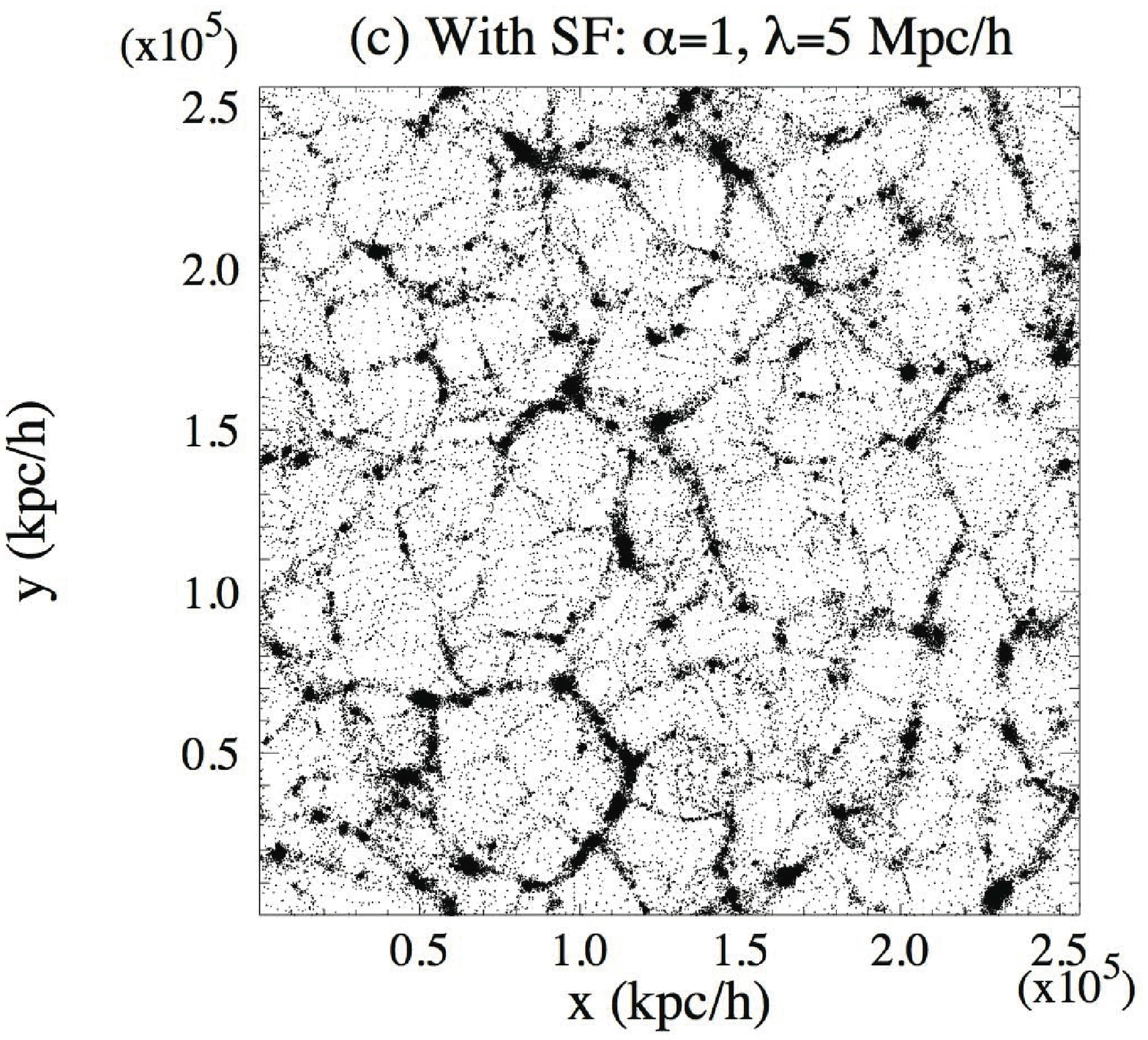}
\caption{(b)--(c) LSS formation process with the SF contribution.
(b) $\alpha=1$ and $\lambda=1$ Mpc$/h$. 
(c) $\alpha=1$ and $\lambda=5$ Mpc$/h$. 
}
\label{fig:LSSF}
\end{figure}

\setcounter{figure}{2}
\begin{figure}
\includegraphics[width=3.5in]{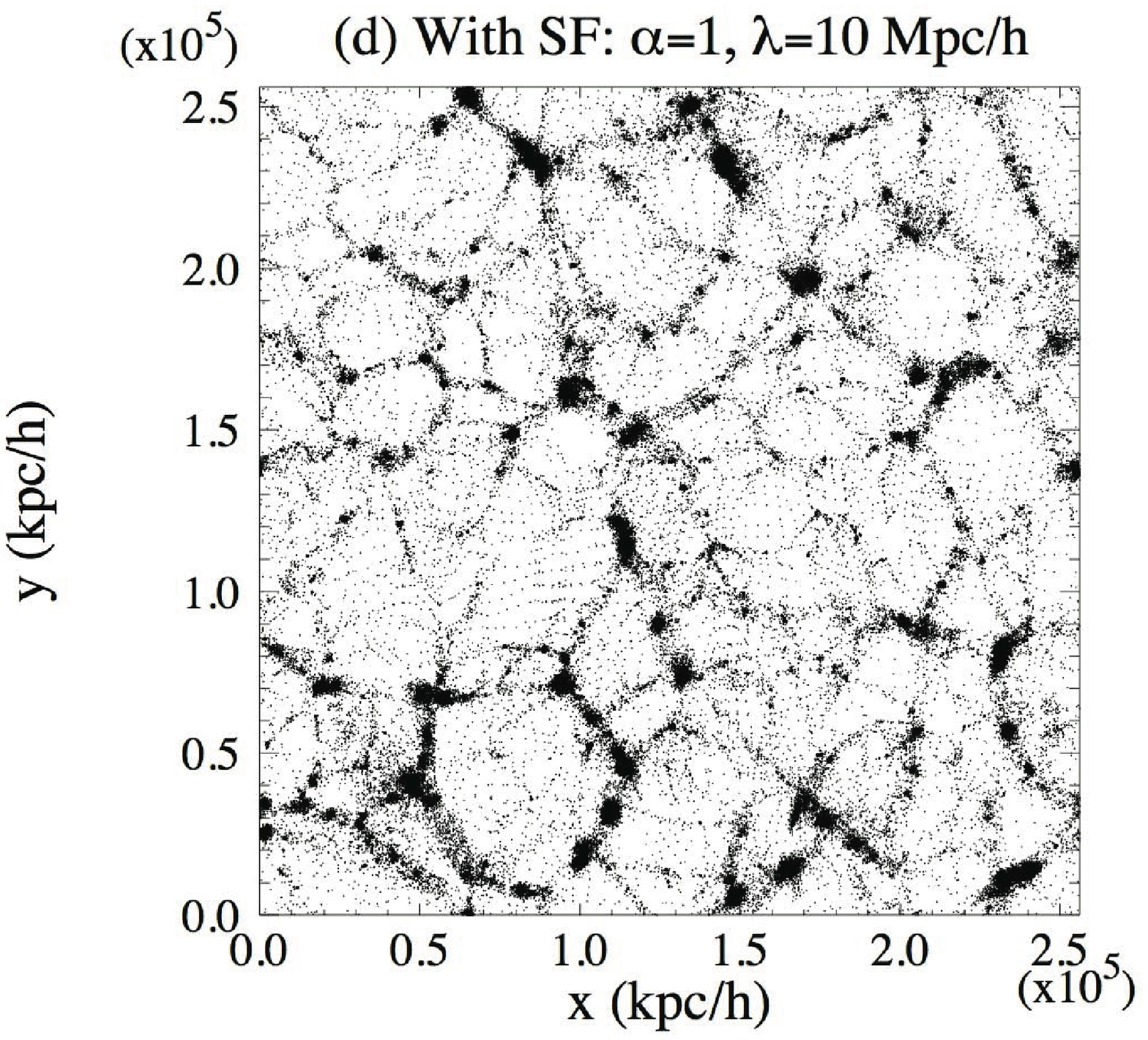}
\includegraphics[width=3.5in]{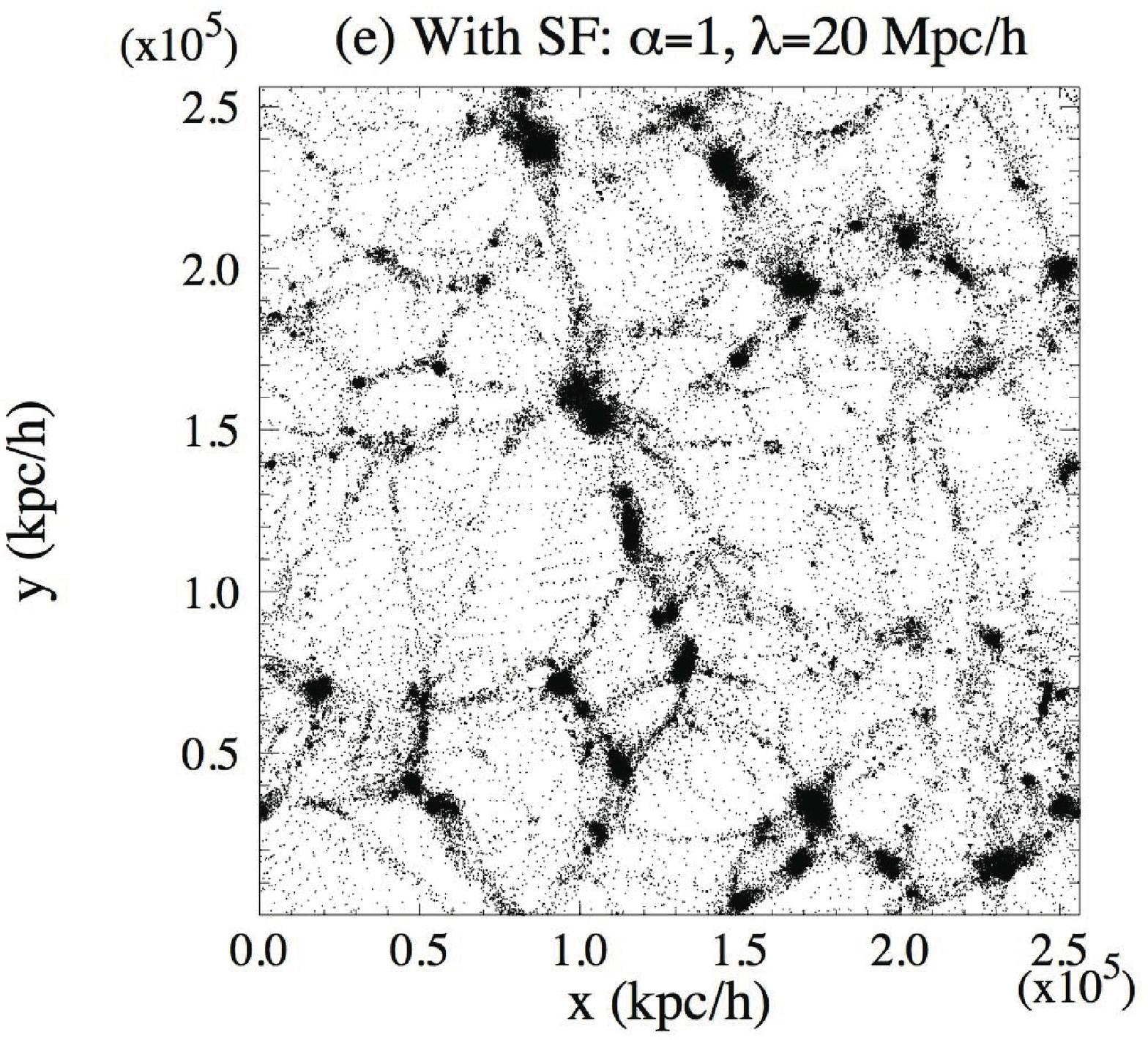}
\caption{(d)--(e) LSS formation process with the SF contribution. 
(d) $\alpha=1$ and $\lambda=10$ Mpc$/h$. 
(e) $\alpha=1$ and $\lambda=20$ Mpc$/h$. 
}
\label{fig:LSSF}
\end{figure}

\setcounter{figure}{2}
\begin{figure}
\includegraphics[width=3.5in]{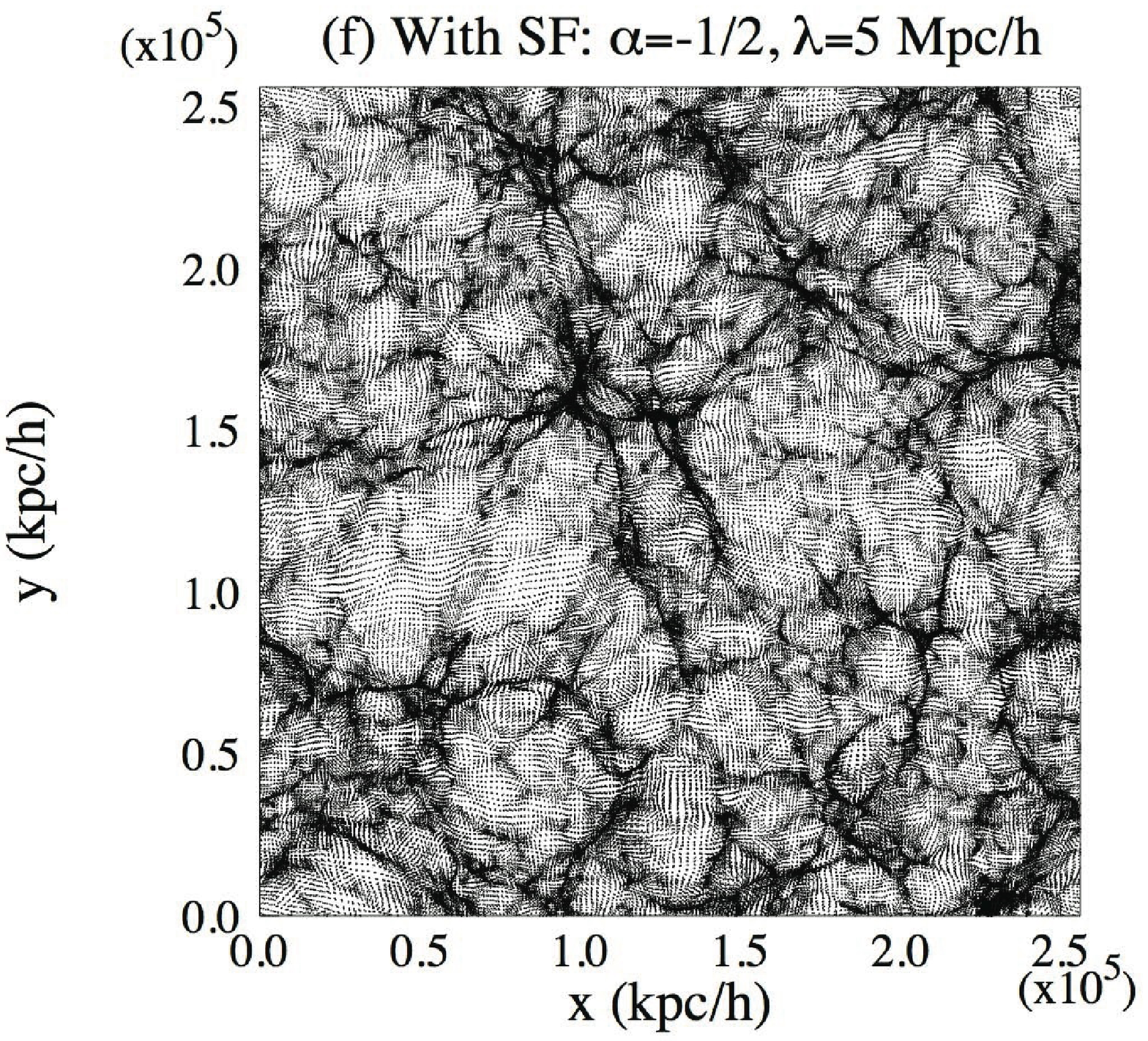}
\includegraphics[width=3.5in]{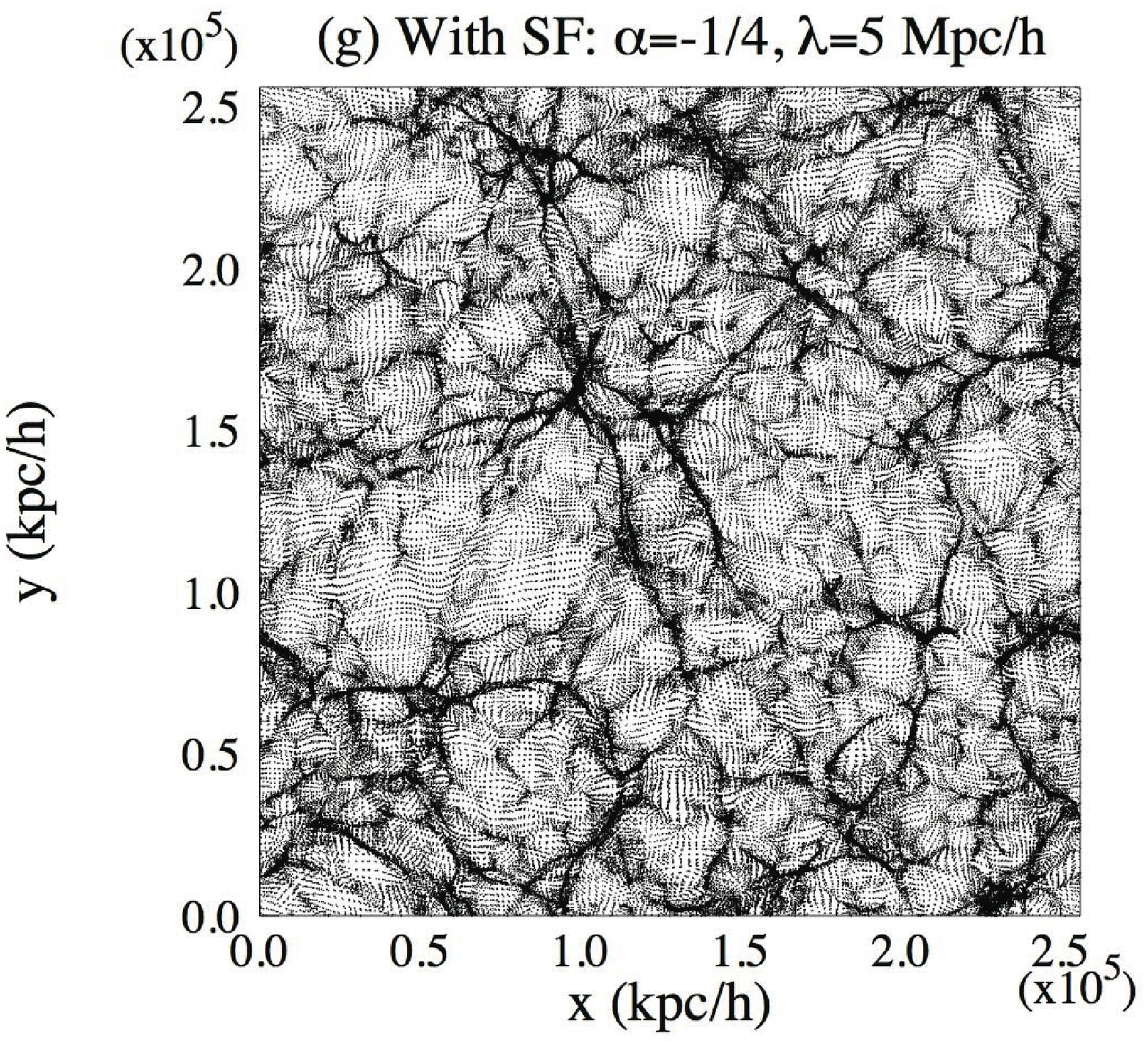}
\caption{(f)--(g) LSS formation process with the SF contribution. 
(f) $\alpha=-1/2$ and $\lambda=5$ Mpc$/h$. 
(g) $\alpha=-1/4$ and $\lambda=5$ Mpc$/h$. 
}
\label{fig:LSSF}
\end{figure}
In Fig. \ref{fig:LSSF_a} we show how the above initial conditions evolve and give us the LSS formation process
without SF and with SF.
In (a) without SF.
In (b) 
with SF: $\alpha=1$ and $\lambda= 1$ Mpc$/h$.
In (c) 
with SF: $\alpha=1$ and $\lambda= 5$ Mpc$/h$.
In (d) 
with SF: $\alpha=1$ and $\lambda= 10$ Mpc$/h$.
In (e) 
with SF: $\alpha=1$ and $\lambda= 20$ Mpc$/h$.
In (f) 
With SF: $\alpha=-1/2$ and $\lambda= 5$ Mpc$/h$.
In (g) 
With SF: $\alpha=-1/4$ and $\lambda= 5$ Mpc$/h$.
%\begin{figure}
%\includegraphics[width=4.5in]{small_box.png}
%\caption{LSS formation process with and without SF. See text for details}
%\label{fig:LSSF}
%\end{figure}

To study the structure formation in the universe we follow the evolution of
the overdensity \cite{Binney},
\begin{displaymath}
\delta(\mathbf{x}) \equiv \frac{\rho(\mathbf{x})}{\rho_0}-1
\end{displaymath}
where $\rho_0$ is the average density over a volume $V$ and $\mathbf{x}$ is the comoving distance related to the physical density by $\mathbf{r}=a(t)\mathbf{x}$.
In the linear regime $\delta \ll 1$.

 The correlation function tell us how $\delta$ is correlated in two nearby
points $\mathbf{x}'$ and $\mathbf{x}'+\mathbf{x}$,
\begin{displaymath}
\xi(\mathbf{x}) \equiv \langle \delta(\mathbf{x}')  \delta(\mathbf{x}'+\mathbf{x}) \rangle
\end{displaymath}
and the power spectrum is the Fourier transform of the correlation function,
\begin{displaymath}
\xi(x)= \frac{1}{V} \sum_\mathbf{k} P(\mathbf{k}) e^{i\mathbf{k}\cdot\mathbf{x}}
\end{displaymath}

\begin{figure}
\begin{center}
\includegraphics[width=3.8in]{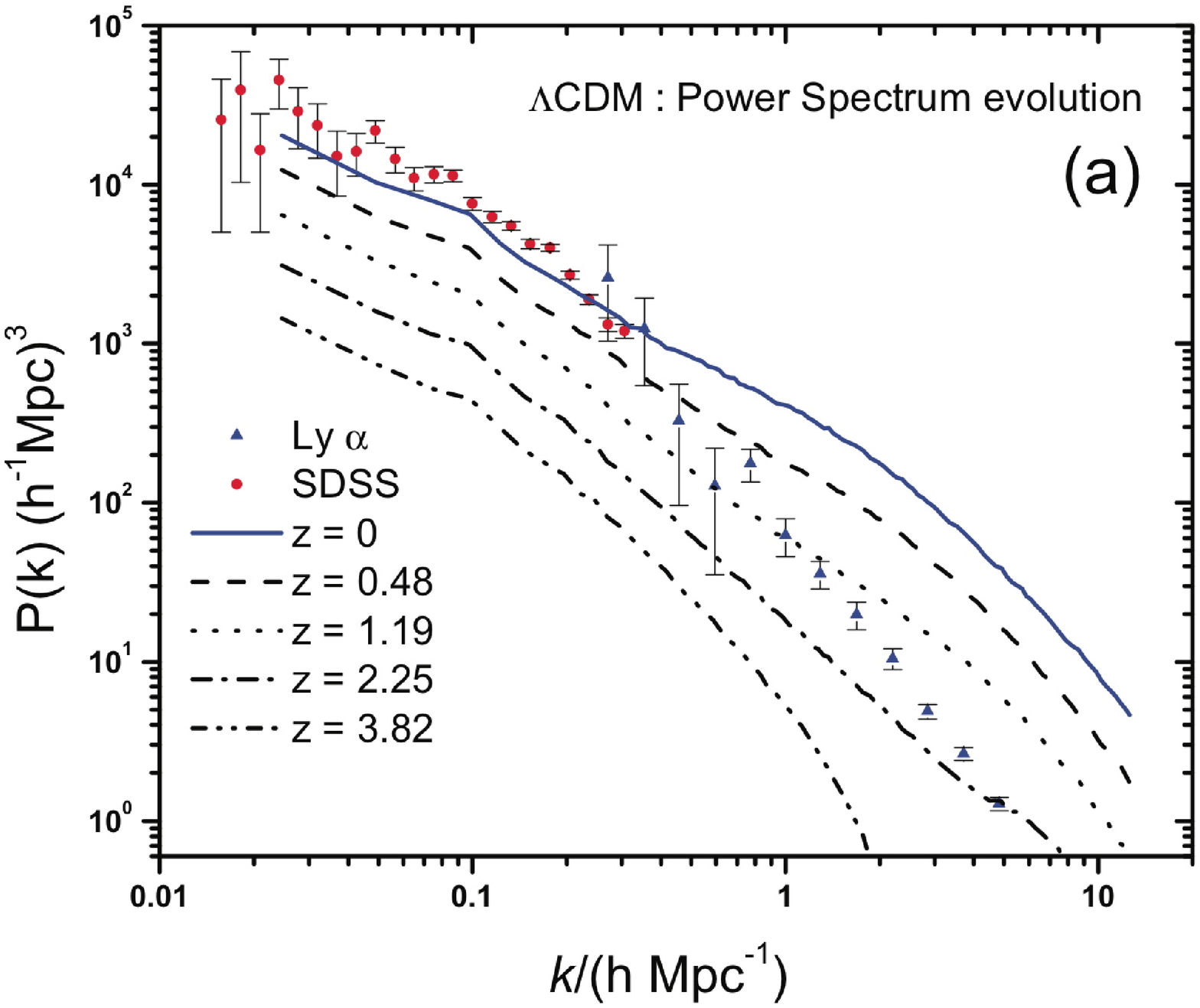}
\includegraphics[width=3.8in]{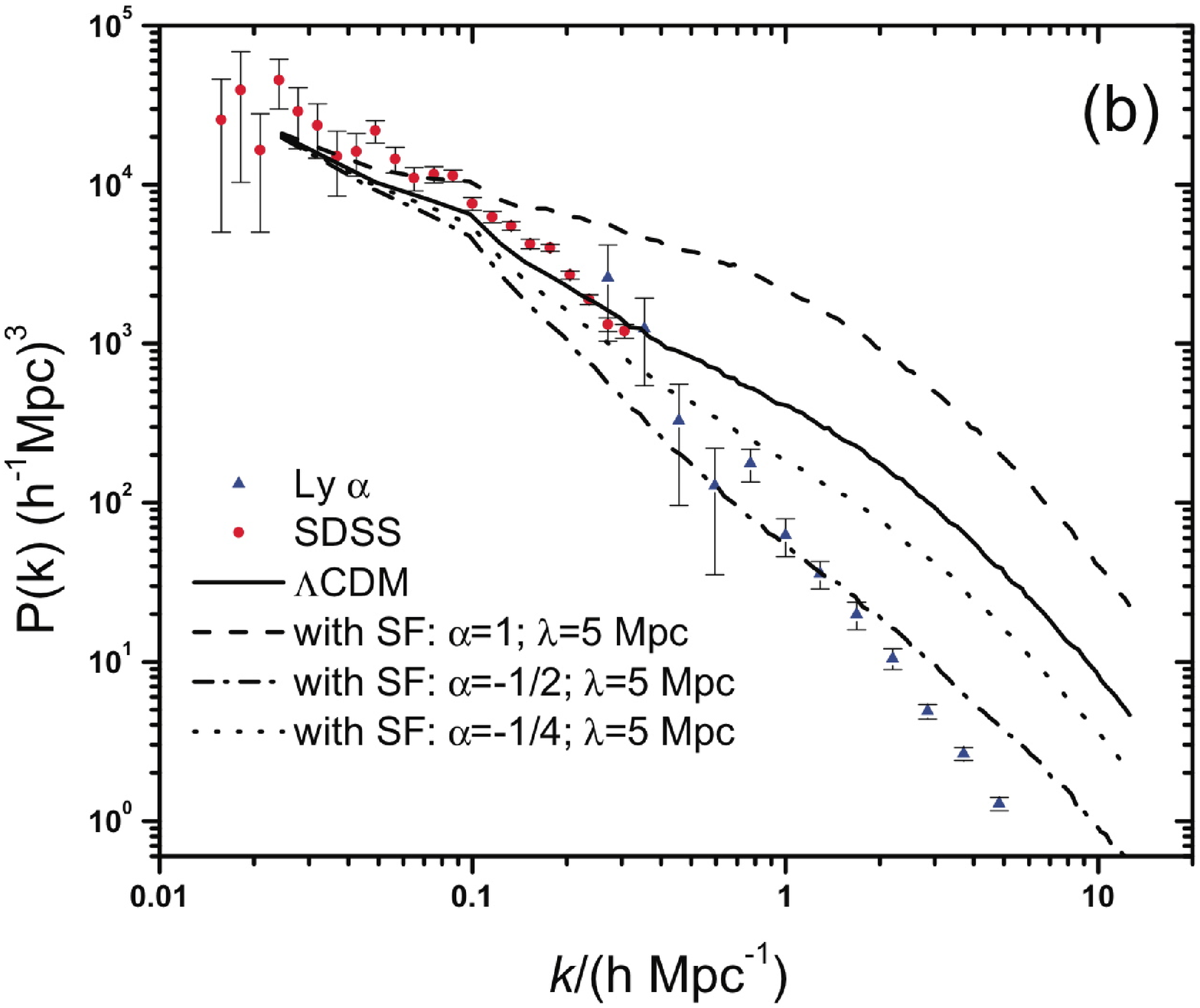}
\end{center}
\caption{(a) Evolution of the power spectrum for the case which there is no SF.
(b) The power spectrum including SF for several values of parameter $\alpha$
and with $\lambda=5$ Mpc$/h$. Experimental data are from galaxies in the Sloan Digital Sky Survey 
(SDSS) catalog and from an analysis of the Lyman-$\alpha$ forest\cite{Tegmark2004}}\label{fig:powmes}
\end{figure}

In Fig.  \ref{fig:powmes} (a) we show the 
evolution of the power spectrum for the 
big box $L=256$ Mpc$/h$ without SF,  for several values of the redshift $z$. 
We have used the \textsc{POWMES} code to compute the matter power spectrum \cite{Powmes2009}.
In this figure we can appreciate how it is forming structures, upper curve of
the power spectrum for the present epoch, $z=0$.
 Fig.  \ref{fig:powmes}  (b) 
 shows the power spectrum for the same values as in Fig.  \ref{fig:LSSF_a}. Continuous line
 is without SF. 
 Upper curve (dashed line) is with SF: $\alpha=1$ and $\lambda=5$ Mpc$/h$.
Lower curve (dashed-dotted line) is with SF: $\alpha=-1/2$ and $\lambda=5$ Mpc$/h$.
The curve that is just below the continuous line (dotted line) is with SF: $\alpha=-1/4$ 
and $\lambda=5$ Mpc$/h$. More greater values of the power spectrum means more
structure formation. Therefore, the inclusion of a SF modifies the structure formation process.
Depending of the values of its parameters, $\alpha$ and $\lambda$ we can obtain more structure
or less structure at the present epoch, $z=0$.

In our results as shown in Figs. \ref{fig:LSSF_a} (c), (f) and (g), and  \ref{fig:powmes} 
we have used a fix value of $\lambda=5$ Mpc$/h$.
For a given $\lambda$ the role of SF parameter $\alpha$ on the structure formation can be
inferred by looking at equations (\ref{eq_motion}) and (\ref{eq:FSFFactor}) and 
Fig. \ref{fig:FSF} where
we show the behavior of $F_{SF}$ as a function of distance for several values of $\alpha$.
The factor $F_{SF}$ augments (diminishes) for positive (negative) values of $\alpha$ for
small distances compared to $\lambda$, resulting in more (less) structure formation for positive
(negative) values of $\alpha$ compared to the $\Lambda$CDM model. In the case of the
upper curve in Fig. \ref{fig:powmes} (b), for $r\ll \lambda$,  the effective gravitational pull has been
augmented  by a factor of 2, in contrast to the case shown with the lower line 
in Fig. \ref{fig:powmes} (b) 
where it has been diminished by a factor of 1/2. That is why we observe for $r < \lambda$
more structure formation in the case of the upper curve 
in Fig. \ref{fig:powmes} (b) and lesser in the case
of the lower curve in the same figure. The effect is then, for a increasing positive $\alpha$, to
speed up the growth of perturbations, then the halos, and then the clusters, whereas negative
values of $\alpha$ $(\alpha \rightarrow -1)$ tend to slow down the growth. 
We also observe that for the large scale regimen of our simulations ($k < 0.1$ $h/$Mpc)
they tend to predict almost the same structure formation.
From comparison with the experimental results, we see that the $\Lambda$CDM agrees well with
SDSS observations, but predicts more structure formation than observations show in the Lyman-$\alpha$ forest
power spectrum. In general the more favored model is the model with SF with $\alpha=-1/2$ and $\lambda=5$ Mpc$/h$.

%%%%%%%%%%%%%%%% CONCLUSIONES %%%%%%%%%%%%%%%%
\section{Conclusions and final comments}\label{sec:Conclusions}

We have used a general, static STT, that is compatible with local observations by the
appropriate definition of the background field constant, i. e., 
$\langle \phi\rangle = (1+\alpha)/G_N$, to study the LSS formation process. The
initial condition for the several cases (different values of parameter $\alpha$)
was built in such a way that the geometry of the model universe were flat.
Quantitatively, this demands that our models have $\Omega/(1+\alpha)=1$
and this changes the amount of dark matter and energy of the models in
order to have a flat cosmology.

Using the resulting modified dynamical equations, we have studied the LSS formation
process of a $\Lambda$CDM universe. We varied the amplitude and sign of the
strength of the SF (parameter $\alpha$) in the interval $(-1,1)$ and performed
several 3D-simulations with the same initial conditions. From our simulations
we have found that the inclusion of the SF changes the local dynamical properties
of the most massive groups, however, the overall structure is very
similar, as can be seen in Fig. \ref{fig:LSSF}.

The general gravitational effect is that the interaction between dark matter particles
given by the potential $\Phi_N$ (see equation (\ref{eq:pointMassPotential})) changes
by a factor $F_{SF}$, equation (\ref{eq:FSFFactor}), in comparison with the purely
Newtonian case. Thus, for $\alpha > 0$ the growth of structures speeds up in comparison
with the Newtonian case (without SF). For the $\alpha <0$ case the effect is to diminish
the formation of structures.

It is important to note that particles in our model are gravitating particles and that
the SF acts as a mechanism that modifies gravity. The effective mass of the SF
 ($m_{SF}=1/\lambda$) only sets an interaction length scale for the Yukawa term.
 
 In this work we only varied the amplitude of the SF --parameter $\alpha$-- leaving
 the scale length, $\lambda$, of the SF unchanged. However, in other studies we have
 done \citep{mar2009b} we have found that increasing $\lambda$ enhances the structure
 formation process for $\alpha$ positive, and decreasing $\lambda$ makes the
 structures grow at a slower rate. 

We have computed the mass power
spectrum in order to study the LSS formation process.
The theoretical
scheme we have used 
is compatible with local observations because we have defined
the background field constant 
$<\phi>  =  G_{N}^{-1} (1+\alpha)$ or equivalently that the local gravitational constant is 
given by 
$ (1 + \alpha) \langle \phi \rangle^{-1} $, 
instead of being given by $\langle \phi \rangle^{-1}$. 
 A direct consequence of  
the approach is that  the amount of matter (energy) has to be increased 
for positive values of $\alpha$ and diminished  for negative values of $\alpha$ 
with respect to the standard $\Lambda$CDM model 
in order to have a flat cosmological model. 
Quantitatively, our model demands to 
have $\Omega/ (1+\alpha) =1$ and this changes the amount of dark matter and 
energy of the model for a flat cosmological model, as assumed.   
The general gravitational effect is that  the interaction including  the SF changes by a factor 
$F_{SF}(r,\alpha,\lambda) \approx 1+\alpha \, \left( 1+\frac{r}{\lambda} \right)$ for $r<\lambda$ in 
comparison with the Newtonian case. Thus, for $\alpha >0$ the growth of structures speeds up  
in comparison with the Newtonian case.  For the   $\alpha <0$ case the effect is to diminish 
the formation of structures.  For $r> \lambda$ the dynamics is essentially Newtonian.

Comparison with the power spectrums from galaxies in the SDSS catalog and that inferred from
Lyman-$\alpha$ forest observations tell us that $\Lambda$CDM predicts more structure formation
in the regime of smaller scales ($k>0.4$ $h/$Mpc). Whereas,  the model with SF with $\alpha=-1/2$
and $\lambda=5$ Mpc$/h$, follows the general trend of the observed power spectrum. In this way
we are able to construct a model that predicts the observed structure formation in the regime of small scales,
with lesser number of halo satellites than the $\Lambda$CDM model.

%
%\begin{acknowledgement}
%This work received financial support by 
%CONACYT grant number \ CB-2007-01-84133.
%\end{acknowledgement}

%%%%%%%%%%%%%%%% REFERENCIAS %%%%%%%%%%%%%%%%

%\section{Referencias}

\end{document}